\documentclass{cernyrep}

\usepackage{mathtools}

\usepackage{eso-pic}

\newcommand{\mysection}[1]{\section{#1}\label{#1}}
\newcommand{\mysubsection}[1]{\subsection{#1}\label{#1}}
\newcommand{\mysubsubsection}[1]{\subsubsection{#1}\label{#1}}
\newcommand\CoM{CM}
\newcommand\pN{p}
\newcommand\mN{M}
\newcommand\mA{M_A}
\newcommand\mpv{v_\mathrm{p}}

\newcommand\plab{p_{\rm lab}}
\newcommand\avgv{\overline{v}}

\newcommand\hl{\it}
\newcommand\Ignore[1]{ }
\newcommand\bb{{\tt bb}}
\newcommand\be{{\tt be}}
\newcommand\eb{{\tt eb}}
\newcommand\ee{{\tt ee}}
\newcommand\press{P}

\newcommand{\boo}[1]{\tilde{#1}}
\newcommand{\href}[2]{\\ \url{#1}}

\def\MM{M_{\rm mol}}
\def\third{\frac{1}{3}}
\def\hca{\alpha}            % half crossing angle a
\def\cossqa{\cos^2\!\hca}   % cos^2(a)
\def\ovlap{\Omega}

\def\hbarc{\hbar c}

\def\hbarc{\hbar c}
\def\gep{G_{E,\mathrm{p}}}
\def\gmp{G_{M,\mathrm{p}}}

\def\dcos{\mathrm{d}\!\cos}

\def\csq{\cos^2\frac{\theta}{2}}
\def\ssq{\sin^2 {\theta} / {2}}
\def\ssqsq{\sin^4\frac{\theta}{2}}
\def\tsq{\tan^2\frac{\theta}{2}}

\def\tthick{\Theta}

            \usepackage{varwidth}
\usepackage{xcolor}

\usepackage{amssymb}
\usepackage[T1]{fontenc}
\usepackage[bookmarks, colorlinks=true, linktoc=page, pdftex, linkcolor=black, citecolor=black, urlcolor=blue]{hyperref}
\usepackage{fancyhdr}
\fancyhfoffset{4 mm}
\fancypagestyle{ARTTITLE}{%
\fancyhf{} % clear all header and footer fields
\chead{\small Proceedings of the 2017 CERN--Accelerator--School course on \it Vacuum for Particle Accelerators, Glumsl\"ov, (Sweden)} 
\lfoot{Available online at \url{https://cas.web.cern.ch/previous-schools}}
\rfoot{\thepage\hspace*{3mm}}

}
\begin{document}

%\AddToShipoutPicture{\BackgroundPic}

\title{Beam--Gas Interactions}
\author{M. Ferro-Luzzi}
\institute{European Organization for Nuclear Research, Geneva, Switzerland}

\begin{abstract}
An overview of beam--gas interactions in particle accelerators is presented.
The paper is focused on a few examples and tries to present basic concepts with
simple formulae that should allow the reader to perform order-of-magnitude
estimates of beam--gas rates.
\end{abstract}


\maketitle

\thispagestyle{ARTTITLE}

\mysection{Introduction}

In this paper, an overview is given of the role of beam--gas interaction phenomena
in particle accelerators with associated particle physics experiments.
The discussion focuses on the energy range between a few gigaelectronvolts and a few teraelectronvolts.
Covering all phenomena in all types of accelerators is not the purpose
of the paper. Rather, a few examples are covered  and
the reader is referred to the literature for a more thorough discussion.
The origin of the residual gas in an accelerator is not explained here.
It is treated in other lectures of this school \cite{Lec:Fundamentals,Lec:Outgassing,Lec:Dynvac}.
Here, it will just be assumed that some residual gas is present, at some level.

Since the beam pipe vacuum is never perfect, every accelerator design
must address effects due to beam--gas interactions.
Generally, beam--gas interactions are rather seen as a nuisance.
This is due to some undeniable detrimental effects, such as reduction
of the beam lifetime, radiation damage to surrounding equipment
or background signals in particle-sensitive detectors (used either
for beam diagnostics or for physics experiments).
Yet, beam--gas interactions can also be an asset.
They can be used for extensive beam monitoring (beam size, position, slopes,
bunch populations, \etc) and, when gas is injected in a controlled manner
to form an internal target, beam--gas interactions are also used for
physics measurements.
Indeed, in this paper, the gas traversed by the beam will sometimes
be called the `target'.
%       and Ref. \cite{AliceNote}
%   Apart from luminosity, very similar to radiation from collimation.
%   On can calculate that the rates from beam-gas interactions are negligible to
%   those from beam-beam collisions in the  ATLAS, CMS and LHCb  experiments.
%
%  Other exotic accelerators: muons, pions, ions, ... you dream i, were not covered.
This dual view, nuisance versus asset, will be further developed
throughout this paper.

This manuscript is organized as follows.
In Section \ref{The nature of beam--gas interactions},
the general nature of beam--gas interactions is discussed.
Some cross-section formulae and numerical examples for beam--gas interaction
processes are given in Section \ref{Beam--gas interaction cross-sections}.
The losses due to beam--gas interactions and their effect on the beam lifetime
are treated in Section \ref{Beam--gas losses and beam lifetime}.
In Section \ref{Detector background},
 the problem of detector background due to beam--gas interactions
is briefly addressed, while the possibility of imaging the beams with these interactions
is presented in Section \ref{Beam--gas imaging} and the opportunity to make physics experiments
with internal gas targets is outlined in Section~\ref{Gaseous fixed targets}.
Section \ref{Summary and outlook} gives a summary and outlook.

For the sake of conciseness, some  topics have been intentionally left out, notably
radiation from beam--gas interactions, which can be important for the installed instrumentation,
see the lecture on radiation in this school \cite{Lec:Radiation}.

\mysection{The nature of beam--gas interactions}

Beam--gas interactions unavoidably occur when beam particles traverse a region
containing gas of a given density.
The rate of interaction will depend on the properties of the beam and of the target.
The beam is characterized by its energy and particle type. % $E_{\rm beam}$.
Most beams are composed of either electrons, positrons, protons, antiprotons, or ions.
The target is characterized by the gas density and by its nature (which
molecules compose the target).

Let us first consider a gas composed of a single type of molecule.
Each molecule contains the same atoms and the mass of the molecule $\MM$
is approximately given by the sum of the mass numbers of its constituting atoms
$\MM  \approx \sum_i \alpha_i \, M_i$, where $M_i$ is the mass of atom type $i$ and
$\alpha_i$ its multiplicity in the molecule.
For example, for carbon dioxide one would have $M_{{\rm CO}_2} \approx M_{\rm C} + 2\, M_{\rm O}$.
Electron masses and their binding energies are neglected.
The mass of an atom with mass number $A$ is dominated by the nuclear mass, which we call $\mA$.
In this lecture, differences between neutron and proton masses
can be ignored and an approximate nucleon mass
$\mN = 1.67 \times 10^{-27}\Ukg$ shall be used throughout.
Nucleon binding energies in nuclei are also neglected and
the approximation for the mass of a nucleus $\mA \approx A\cdot \mN$
is adopted.
For carbon dioxide, one ends up with $M_{{\rm CO}_2} \approx (12 + 2 \times 16)\, \mN = 44\, \mN$.

The distribution $f(v)\mathrm{d}v$ of the velocity $v$ of atoms in a gas at temperature $T$ is a
Maxwell--Boltzmann distribution:
\begin{equation}
  f(v)\mathrm{d}v = \left(\frac{\MM}{2\pi k_\mathrm{B} T}\right)^\frac{3}{2}  4\pi v^2  \mathrm{e}^{-{\MM v^2}/{2k_\mathrm{B} T}} \mathrm{d}v
         = \left(\frac{1}{\pi}\right)^\frac{3}{2}\,  4\pi\, \xi^2 \mathrm{e}^{-\xi^2}\mathrm{d}\xi ~,
\end{equation}
where $\xi = v/\mpv$ and  $\mpv$ is the most probable velocity\footnote{Obtained by setting $\mathrm{d}f/\mathrm{d}v = 0$.}
  $\mpv = \sqrt{2\, k_\mathrm{B}\, T/\MM}$,
with $k_\mathrm{B} \approx 1.38 \times 10^{-23}~{\rm J/K}$ the Boltzmann constant.
The mean velocity is $\avgv = 2\,\mpv/\sqrt{\pi}$ and
the mean kinetic energy is  $E_{\rm kin} = (3/2)\,k_\mathrm{B} T$.
%\begin{equation}
%   \avgv = \sqrt{\dyfr{8\, k_B\, T}{\pi\, \MM}} = \dyfr{2}{\sqrt{\pi}} \, \mpv \approx 1.28\, \mpv.
%\end{equation}
It is instructive to calculate the mean velocity and mean kinetic energy
%and corresponding momentum $\avgp = \MM\, \avgv$
for some typical cases:
\begin{equation}
  \begin{array}{lrrrrl}
      \mbox{Molecule}&A &\mbox{Mass}& T         &  \avgv                    & E_{\rm kin}          \\%& \avgp
   \mbox{H}_2      & 1  & 2\mN  &  10~{\rm K}   &  \approx  320~\mbox{m/s}  & \approx \phantom{0}1.3~{\rm meV}\\%& \approx 2~{\rm keV}/c
   \mbox{H}_2      & 1  & 2\mN  & 300~{\rm K}   &  \approx 1780~\mbox{m/s}  & \approx 39~{\rm meV} \\%& \approx 11~{\rm keV}/c
   \mbox{Ar}       & 40 & 40\mN & 300~{\rm K}   &  \approx  400~\mbox{m/s}  & \approx 39~{\rm meV} \\%& \approx 50~{\rm keV}/c
  \end{array}
\end{equation}

In the laboratory frame, the beam particle velocity is highly directional
(along the beam axis) and approaches the speed of light $c =3 \times 10^8~{\rm m/s}$,
while molecules have a thermal velocity in the range of 100 to a few 1000~m/s,
generally pointing in a random direction.
Therefore,
%the energy, or velocity, of the target gas is in general negligible.
%negligible when dealing with beams with energy of more than 1~GeV (or 1~GeV per nucleon).
when considering beam--gas interactions, the momenta of the rest gas molecules
and their constituents %, when compared to beam particle momenta in the range of interest,
can be safely neglected. They will be considered at rest (zero momentum),
as if `frozen in space', when beam particles traverse the target.
These generic features are summarized in Table \ref{tab:beam_vs_gas}.

\begin{table}
\centering
\caption{Some key features of beams and residual gas}\label{tab:beam_vs_gas}
\begin{tabular}{llll}\hline\hline
              & {Beam}  &  & {Residual gas} \\\hline
   Particles & $p^\pm$, $e^\pm$, $^{208}$Pb$^{82+}$, \dots  & & Molecules (H$_2$, CH$_4$, CO, \dots) \\
   Velocity &  $\approx c = 3 \times 10^8~{\rm m/s}$ & & $\approx 100 \dots
   1000~$m/s ($\ll c$)\\
   Energy   & typically MeV to TeV %, and often $E_{\rm beam} \gg mc^2$
             & &  Thermal, $E_{\rm kin}  \approx 1$--$100$\,meV \\\hline\hline
\end{tabular}
\end{table}

Beam particles can interact with residual gas molecules through two types of force:
\begin{itemize}
    \item The {\hl strong interaction} (or `hadronic' interaction):
          this is relevant only for {\hl hadron} beams (protons, ions, \etc),
          which interact with the {\hl nuclei} of the residual gas molecules.
          The strong interaction is \dots strong! But its range is short, about
          $10^{-15}~{\rm m}$ or $1~{\rm fm}$, which is approximately the size of a nucleon.
    \item The {\hl electromagnetic interaction}:
        this force is relevant for all accelerated beams. (Indeed, acceler\-ation makes use
         of the electromagnetic interaction!) The beam can interact with
both
          {\hl atomic nuclei} and {\hl atomic electrons} of the residual gas.
          The strength of the electromagnetic interaction is typically
          reduced by the factor $\alpha \approx 1/137$ compared with the strong
          interaction. But the range is much longer (infinite!).
\end{itemize}
The weak interaction and gravitation are irrelevant in the context of this paper.

Generally, beam--gas interactions are more relevant for cyclical accelerators
than for linear accelerators, since the beam particles pass through
the gas several times and one has to worry about the lifetime of the beam.
Let us consider a bunch of beam particles, with population $N$, traversing
a region containing gas.
Denoting by $\rho(z)$ the density of gas molecules along the beam path $z$
and by $\tthick =  \int\rho(z)\mathrm{d}z$  the {\hl target thickness} (or areal density),
then the probability $\mu$ for a given interaction process to occur per bunch and per pass
is
  \begin{equation}
    \mu = \sigma_{\rm proc} \cdot N\cdot \tthick~.
  \end{equation}
The proportionality factor $\sigma_{\rm proc}$ is the {\hl cross-section}
of the  given interaction process.
The units of $\sigma_{\rm proc}$ are those of surface area.
Since the cross-sections involved are tiny quantities,
the unit `barn' (symbol `b') is introduced:
 $1\,{\rm b} = 10^{-24}\,{\rm cm}^2$.

Repeating the beam passage many times, say, at a frequency $f$,
produces a rate of interactions $R$ given by
      \begin{equation}
        R = f \cdot \mu = \sigma_{\rm proc} \cdot L \, ,
        \label{eq:rate1}
      \end{equation}
where $L$ is the instantaneous luminosity
      \begin{equation}
        L =  f\cdot N \cdot \tthick~,
        \label{eq:rate2}
      \end{equation}
a measure of how intense or dense the beam and target are.
Numerical examples will appear in Section \ref{Beam--gas imaging}.

\begin{figure}
      \includegraphics[width=0.98\textwidth]{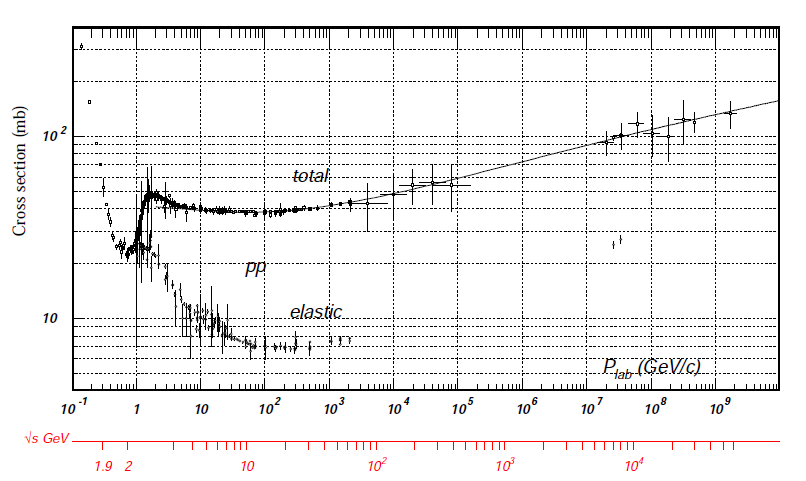}
      \caption{Proton--proton cross-section; figure taken from Ref.~\protect~\protect\cite{PDG}}
      \label{fig:pp-cross-section}
  \end{figure}

Figure~\ref{fig:pp-cross-section}, taken from Ref.~\cite{PDG}, shows the measured hadronic cross-section of
proton--proton col\-lisions, p + p. Both the total and elastic hadronic cross-sections are shown.
The cross-section is shown for both the fixed-target type of experiment (with $\plab$
being the momentum of the beam particles) and the equivalent collider experiment,
in which the colliding protons have the same (but opposite) momenta.
In this latter case, the quantity $\sqrt{s}$ is used to express the total energy of the
colliding system.
Indeed, changing the frame of reference, \ie moving the observer
by a constant speed relative to the gas and beam (which changes
the apparent speed of the gas and beam particles)
does not change the observed number of interactions (all observers should
count the same number of interactions!).
A brief digression on special relativity is needed.

Formally, the change of reference frame is a `Lorentz boost'.
In special relativity, the length squared $r^2$ of a four-vector $r$ is defined as
\begin{equation}
   \left.
   \begin{matrix*}[l]
   r_0   \\
   \mathbf{r}=\begin{pmatrix}r_1\\r_2\\r_3\end{pmatrix}
   \end{matrix*}
   \quad \right\}
   \quad\quad
   \Rightarrow
   \quad\quad
   r=\begin{pmatrix}r_0\\r_1\\r_2\\r_3\end{pmatrix}
   \quad\quad
   \Rightarrow
   \quad\quad
   r^2 \equiv r_0^2 - r_1^2 - r_2^2- r_3^2 = r_0^2 - \mathbf{r}^2 ~,
\end{equation}
where $\mathbf{r}^2 = \mathbf{r} \cdot \mathbf{r}$ is the usual three-dimensional scalar product.
Here, $r_0$ and $\mathbf{r}$ can be time and position intervals,
energy and momentum, charge and current densities, \etc
As we will see next in a concrete example, the four-vector length is an invariant,
\ie it does not depend on the observer's velocity % relative to the particle.
(it is frame-independent), while the individual components $r_i$, or even
the classical three-vector length $|\mathbf{r}|$, are not.

For a particle of rest mass $m$, with energy $E$ and momentum $\mathbf{p}$,
$E$ and $\mathbf{p}$ form together the four-momentum of the particle, here
written $cp$:
\begin{equation}
   \left.
   \begin{matrix*}[l]
   E   \\
   \mathbf{p}=\begin{pmatrix}p_x\\p_y\\p_z\end{pmatrix}
   \end{matrix*}
   \quad \right\}
   \quad\quad
   \Rightarrow
   \quad\quad
   cp= \begin{pmatrix}E\\cp_x\\cp_y\\cp_z\end{pmatrix}
\end{equation}
where $c$ is the speed of light. (In most textbooks, $c$ is set to one and removed from equations. Equation~\protect\ref{eq:fourmom}
  would read $p^2 = E^2 - \mathbf{p}^2$ and Eq.~\protect\ref{eq:restmass}
  would read $\sqrt{p^2} = \hat{E} = m$.
  Here, we keep $c$ explicit, to avoid confusing anyone with specialized units.)
Therefore, the four-momentum's length squared is
\begin{equation}
   c^2p^2 \equiv E^2 - c^2p_x^2 - c^2p_y^2- c^2p_z^2 = E^2 - c^2\mathbf{p}^2 ~.\label{eq:fourmom}
\end{equation}
The Lorentz boost works as follows.
Consider an observer ${\cal O}$ moving along $z$ with a velocity $v$.
Define $\beta =  {v} / {c}$ and $\gamma = (1-\beta^2)^{- {1} / {2}}$.
Then special relativity tells us that the four-momentum vector $c\boo{p}$ of the particle
as seen from ${\cal O}$'s frame is
\begin{equation}
   c\boo{p} =
   \begin{pmatrix}
       \boo{E}   \\
      c\boo{p}_x \\
      c\boo{p}_y \\
      c\boo{p}_z
   \end{pmatrix} =
  \begin{pmatrix*}[c]
           \gamma & 0 & 0 & -\beta\gamma \\
            0     & 1 & 0 &  0           \\
            0     & 0 & 1 &  0           \\
     -\beta\gamma & 0 & 0 & \gamma \\
  \end{pmatrix*}
   \begin{pmatrix}
       E   \\
      cp_x \\
      cp_y \\
      cp_z
   \end{pmatrix} =
   \begin{pmatrix}
       \gamma\, ( E -\beta\,cp_z) \\
      cp_x \\
      cp_y \\
       \gamma\, (cp_z -\beta\, E)
   \end{pmatrix} ~.\label{eq:boost}
\end{equation}
One easily verifies that the four-vector length squared is not affected by
the boost transformation:
\begin{align}
   c^2\boo{p}^2
    &= (\gamma\, ( E -\beta\,cp_z))^2 - (cp_x)^2 - (cp_y)^2 - (\gamma\, (cp_z -\beta\, E))^2 \\
   %&=& \gamma^2\, ( ( E^2 + \beta^2\,(cp_z)^2 - 2 \,\beta\, E\, cp_z ) - ((cp_z)^2 + \beta^2\, E^2 - 2\,\beta\,E\, cp_z ) ) \\
   %&=& \gamma^2\, ( (1-\beta^2)\, E^2  - (1-\beta^2)\,(cp_z)^2  ) - (cp_x)^2 - (cp_y)^2  \\
    &= E^2  - (cp_z)^2  - (cp_x)^2 - (cp_y)^2  \\
    &= c^2 p^2 ~.
\end{align}
The quantity $\sqrt{c^2p^2}$ is identified with the particle's mass or rest energy, \ie
the total internal energy $\hat{E}$ available in the frame (denoted here by the $\,\hat{ }\,$ symbol)
where the particle does not move (it has zero momentum, $\mathbf{\hat{p}}=0$):
\begin{equation}
   \sqrt{c^2 p^2} = \hat{E} = mc^2~.\label{eq:restmass}
\end{equation}
Indeed, in particle physics, one generally writes particle masses in units of GeV/$c^2$
rather than kilograms.
In these units, the nucleon mass introduced earlier is $\mN = 0.939~{\rm GeV}/c^2$.

In the case of a particle with four-momentum $cp_1 = (E_1,c\mathbf{p}_1)$
colliding with another particle with four-momentum $cp_2 = (E_2,c\mathbf{p}_2)$,
the same invariant quantity (usually called $s$) can be defined
for the sum of the two four-vectors:
\begin{equation}%
   cp = cp_1 + cp_2  =
   \begin{pmatrix}
       E_1 + E_2         \\
      cp_{1,x} + cp_{1,x} \\
      cp_{1,y} + cp_{1,y} \\
      cp_{1,z} + cp_{1,z}
   \end{pmatrix}  \, ,
\end{equation}
which is the total four-momentum of the two-particle system.
The invariant $s$ is just $c^2p^2$:
\begin{align} \label{eq:s}
          s \equiv c^2p ^2 & = (E_1+E_2)^2 - c^2\, ( \mathbf{p}_1 + \mathbf{p}_2 )^2 \\[1mm]
                    & = m_1^2c^4+m_2^2c^4 + 2\, (E_1\, E_2 - c^2\, \mathbf{p}_1 \cdot\mathbf{p}_2 )~.
     \end{align}
%One can easily check that $s$ is the same for any observer frame.
In fact, one sees that $\sqrt{s}$ is the {\hl total available energy} $E_1+E_2$
in the system, where $\mathbf{p}_1 =-\mathbf{p}_2$, \ie $\mathbf{p}_1 +\mathbf{p}_2 = 0$, which is
frequently called the `centre of mass' system ({\CoM} system) or `centre of momentum' system.

In the accelerator physics world, there are some standard configurations for
the beam kinematics:
\begin{itemize}
   \item[a)] The like-particles collider mode, with
   $\mathbf{p}_1 =-\mathbf{p}_2$ and $m_1 = m_2$, therefore $E_1 = E_2 = E$.
   In this configuration, the laboratory system is coincident with the {\CoM} frame.
   Here, one obtains
   \begin{equation}\label{eq:collider}%
       \sqrt{s} = E_1+E_2 = 2\,E~.
   \end{equation}
   \item[b)] The fixed-target mode, with
   $\mathbf{p}_1 \neq 0$ and $\mathbf{p}_2 = 0$. In this case, since $E_2 = m_2c^2$, one has
   \begin{align}\label{eq:fixedtarget}%
          \sqrt{s} & = (m_1^2c^4+m_2^2c^4 +2\, E_1 \, m_2c^2)^\frac{1}{2}  \\[1mm]
               &  \approx (2\, E_1 \, m_2c^2)^\frac{1}{2} \qquad~ \qquad ~~~({\rm if~~}E_1\gg m_1c^2,\,m_2c^2) ~.
     \end{align}
   \item[c)] The asymmetric collider mode,\footnote{For example, the asymmetric $B$
             factories at KEK and SLAC, the HERA collider at DESY, and the LHC when
             colliding $p$ on Pb.} with:
   $\mathbf{p}_1 \neq -\mathbf{p}_2$, usually $\mathbf{p}_1 \cdot\mathbf{p}_2 \approx -|\mathbf{p}_1|\cdot |\mathbf{p}_2|$ and
      sometimes $m_1 \neq m_2$.
   Here, one obtains
   \begin{align}\label{eq:asymcollider}%
       \sqrt{s} &= (m_1^2c^4+m_2^2c^4 + 2\, (E_1\, E_2 + c^2\, |\mathbf{p}_1|\cdot |\mathbf{p}_2|))^\frac{1}{2} \\[1mm]
               &\approx 2\, (E_1\, E_2 )^\frac{1}{2}  \qquad ~~~(\mbox{if~~} E_1,\,E_2\gg m_1c^2,\,m_2c^2)~.
     \end{align}
\end{itemize}

These formulae will be used in the following to obtain cross-sections
for beam--gas interactions (\ie the fixed-target case) at the correct energy by deducing
them from cross-sections, sometimes meas\-
ured in a different mode.
For example, for the LHC with 6.5~TeV beam energy, the {\CoM} energy in the (a) and (b)  modes
becomes:
\begin{itemize}
   \item[a)] p + p collider mode at 6.5~TeV beam energy:  $\sqrt{s} = 13~$TeV;
   \item[b)] p + $^1$H beam--gas at 6.5~TeV beam energy:  $\sqrt{s} \approx 110~$GeV.
%  \item[c)] $p+$$^208$Pb$^{82+}$ collider mode with $5.02~Z$TeV beam energy: $\sqrt{s} \approx 110~$GeV.
\end{itemize}

\mysection{Beam--gas interaction cross-sections}

In this section, some approximate formulae for estimating rates of beam--gas interactions are given.
If one is interested in beam--gas losses or beam lifetimes, the relevant cross-section
is the total cross-section, if one assumes, to first-order approximation, that
any interaction will disturb the beam particle strongly enough that it can be considered
a lost particle. This is not necessarily true, for example, in the case of particles scattered
elastically at very small angles,
but we shall ignore this here.

In what follows, $A$ and $B$ denote nucleon numbers, as well as the corresponding particle species.

\mysubsection{Proton beams}

Let us forget the atomic electrons of the gas atoms, for a moment, and
concentrate on the interaction of a proton beam particle with laboratory momentum
$\plab$ impinging on gas nuclei.
As discussed, the initial thermal momentum of the nuclei is negligible.
In the case of proton beams traversing a gas target (residual gas),
the relevant interaction is the hadronic interaction.
At large impact parameter (larger than about 1~fm), where the
strong interaction becomes inactive, the electromagnetic interaction
becomes visible and will be dominated by elastic scattering.
This is indeed what causes `multiple scattering'  when a charge particle passes
through matter.
However, the effects we are concerned with (detector background, induced radiation,
beam losses, beam imaging, \etc) are dominated by nuclear inelastic beam--gas interactions,
which are generally strong interaction processes producing several outgoing particles.

Take the case of a proton collider operating at the proton beam energy $E$,
and assume that the main residual gas component is hydrogen.
The relevant cross-section for the beam--gas rates is that of a proton beam
impinging on a proton ($^1$H) fixed target ($B = A = 1$).
If the protons in the beam are ultra-relativistic (\ie $c\plab \gg \mN c^2$),
the beam proton momentum in the lab is just $c\plab\approx E$ and the cross-section,
without further ado,
can be read out directly from the graph of Fig.~\ref{fig:pp-cross-section},
which gives both the elastic and total hadronic cross-sections.
The inelastic cross-section is the difference between the two.
For this specific case, p + p, cross-section data are available for a huge range of laboratory
beam momenta, $\plab$.

For other target gases, parametrizations of the inelastic cross-section for p + $A$ scattering
can be found in the literature \cite{Carvalho}.
Owing to the short range of the strong interaction, the interaction process should be dominated by the
impinging proton inelastically interacting with a single nucleon inside the nucleus.
If one naively imagines the nucleus as a blob composed of independent and incompressible
target nucleons,  which form some kind of `dark or solid sphere' of radius $r$, then
the radius of this naive object should roughly scale with the cubic root of the volume,
therefore as $r \propto A^\frac{1}{3}$, where $A$ is the number of nucleons in the nucleus.
Thus, one would expect the p + $A$ inelastic cross-section to vary as
\begin{equation}
    \sigma_{\mathrm{p}+A} \approx \sigma_{\mathrm{p}+\mathrm{p}} \cdot A^{\alpha}
    \, ,
\end{equation}
with an exponent $\alpha$ close to 2/3 (the surface of the sphere's shadow!).
Here,  $\sigma_{\mathrm{p}+\mathrm{p}}$ is, depending on the purpose,
the total or inelastic p + p cross-section at the equivalent
nucleon--nucleon {\CoM} energy, denoted $\sqrt{s_\mathrm{NN}}$.
To calculate  $\sqrt{s_\mathrm{NN}}$, one considers that the impinging proton
interacts principally with a single nucleon inside the nucleus. Then
\begin{equation}
    \sqrt{s_\mathrm{NN}} \approx (2\, \mN c^2 \, c\plab)^\frac{1}{2} ~.
\end{equation}
At the LHC, with $\plab = 6.5~{\rm TeV}/c$, one obtains $\sqrt{s_\mathrm{NN}} = 110~{\rm GeV}/c$.
From Fig.~\ref{fig:pp-cross-section}, one may retrieve $\sigma_{\mathrm{p}+\mathrm{p},{\rm inel}}\approx 40~$mb
and estimate, for example, $\sigma_{\mathrm{p}+{\rm Ne}} \approx 320~$mb.

In reality, the measured exponent for inelastic scattering is closer to $\alpha \approx 0.7$.
A fit of proton--nucleus ($A>6$) inelastic cross-section data for proton beam energies in the range 150--400~GeV
gives, for example (when rescaled to the HERA proton energy of 920~GeV, or $\sqrt{s_\mathrm{NN}} = 41.6~$GeV)~\cite{Carvalho},
\begin{equation}
   \sigma_{\mathrm{p}+A,{\rm inel}} = (43.55\pm0.40)~{\rm mb}~ A^{0.7111\pm0.0011}~.
    \label{eq:sigmapA_inel}
\end{equation}

% CMS :
% Published in Physics Letters B as doi:10.1016/j.physletb.2016.06.027.
% The inelastic cross section is measured to be
% $\sigma_{{\rm inel},p+Pb} = 2061~$mb (with about 3-4\% accuracy). %(2061 \pm 3 (stat) \pm 34 (syst) \pm 72 (lumi) )mb$
% They don't give the $\sqrt{s_{NN}}$. I calculate it  to be $5.02~$TeV from
% $E_p = 4~$TeV and $E_{\rm Pb}=4~Z$TeV, which gives
% $\sqrt{s_{NN}} = 2\cdot 4\sqrt{\dyfr{82}{208}} = 5.02$~TeV.
% The $p+p$ inelastic cross section at $\sqrt{s_{NN}}=41.6~$GeV (HERA) is about 33~mb,
% while it is about 65~mb at $\sqrt{s_{NN}}=5~$TeV.
% Therefore, the coefficient of expression \ref{eq:sigmapA_inel} should be
% rescaled up by a factor 65/33.
% One obtains $\sigma_{{\rm inel},p+Pb} = 3800$~mb. That is a bit too much.
% But not so bad, considering the two orders of magnitude extrapolation
% in $\sqrt{s_{NN}}$!

For more precise estimates of the cross-section, several generator models
of hadronic proton--nucleus collisions and simulation tools  exist, \eg
%Three models are based on the
%Gribov-Regge formalism: DPMJET 3.06~\cite{DPMJET}, EPOS-LHC~\cite{EPOS-LHC},
%and QGSJETII-04~\cite{QGSJETII-04}; and a fourth
%one is based on a minijet+Glauber approach: HIJING 1.383~\cite{HIJING-1.383}.
 DPMJET 3.06~\cite{DPMJET}, EPOS~(Ref. \cite{Epos} and references therein),
 QGSJETII-04~\cite{QGSJETII-04} and HIJING 1.383~\cite{HIJING-1.383},
   FLUKA~\cite{FLUKA},
   GEANT~\cite{Geant}, and
   Pythia~\cite{Pythia}.
%  HIJING  Model,~\cite{Hijing}
   %SixTrack~\cite{Sixtrack}.
%The value of $\sigma_{{\rm inel},p+Pb}$ is compatible with that expected
%from the proton-proton cross section at 5.02~TeV scaled up within a simple Glauber
%approach to account for multiple scatterings in the lead nucleus, indicating that
%further net nuclear corrections are small.

For low-energy beams (10~MeV to a few GeV), a number of cross-sections have been
parametrized, for example in Ref.~\cite{LowEnergy}.

Note also that, at energies $\sqrt{s}$ of a few gigaelectronvolts or more, total and inelastic
hadronic cross-sections for {\hl antiproton}--gas interactions are quite similar to those
for proton--gas interactions (as inferred from $\bar{\mathrm{p}}+\mathrm{p}$ and $\mathrm{p}+\mathrm{p}$ data, see Ref.\cite{PDG}).

\mysubsection{Ion beams}

 In this section, we consider the collisions $A+B$, where $A$ and $B$ are two types of nuclei
 of charge $Z_A$ and $Z_B$ and mass $M_A \approx A  M_N$, $M_B \approx B  M_N$, respectively.
 We assume that $A$ is the beam and $B$ is the fixed target (gas).

 %If the one is a beam, we assume the nuclei are fully stripped, i.e. the charge is
 %that of the nucleus $Z_A$, respectively $Z_B$.
 For a given beam momentum $\plab$ ($= p_A$),
 the momentum $\pN$ carried by each nucleon of nucleus $A$ is approximately given by
 \begin{equation}
    \pN \approx\frac{\plab}{A} ~.
 \end{equation}
 For example, when colliding $^{208}$Pb$^{82+}$ on $^{208}$Pb$^{82+}$ at the LHC
 ($A = B = 208$, $Z_A=Z_B=Z=82$)
 with a beam rigidity $R = p/Z= 5.02~$TeV
 (beam energy $E = 5.02~Z$\,TeV), the typical {\CoM} collision energy of two nucleons is
 \begin{equation}
    \sqrt{s_\mathrm{NN}} \approx 2\, \frac{E}{A} =  5.02~{\rm TeV}~\frac{Z}{A}
                  = 5.02~{\rm TeV}~\frac{82}{208}   %(soit 39,4 %) ~.
                  = 1.98~{\rm TeV}~.
 \end{equation}
 For this same beam, the NN equivalent {\CoM}  energy when $A$ impinges on a rest gas nucleus $B$ is
 \begin{equation}
    \sqrt{s_\mathrm{NN}} \approx (2\, \mN c^2 \, c\pN)^\frac{1}{2}
                  = \left(2\, \mN c^2 \,  \frac{c\plab}{A} \right)^\frac{1}{2}
                 \approx 61~{\rm GeV} ~,
 \end{equation}
 independent of the target type $B$.

As already mentioned, the cross-section does not depend on the observer's velocity.
Therefore, the  cross-section for an ion beam impinging on hydrogen nuclei, \ie protons,
%(molecules composed of two $^1$H atoms)
is easily obtained from the `reverse' case, where the proton beam impinges on the ion at rest,
see Section~\ref{Proton beams}.
For gases other than hydrogen, a naive formula for an order-of-magnitude estimate of the
inelastic cross-section at high energy is
\begin{equation}
   \sigma_{A+B} = \sigma_{\mathrm{p}+\mathrm{p}} \cdot (A^\third + B^\third)^2~.
\end{equation}
Semi-phenomenological formulae exist, which can give somewhat better estimates,
for example~\cite{Hoang}
\begin{equation}
     \sigma_{A+B} = 54~{\rm mb}\cdot \left( A^\third + B^\third - \frac{4.45}{A^\third + B^\third} \right)^2
\end{equation}
at 1.88 GeV/nucleon.

Again, this formulae can be used for `order-of-magnitude' cross-section estimates.
For more precise estimates, one should preferably resort to experimental data or, if the
latter are not available in the energy range of interest or for the desired beam and target species,
one should exploit modern generator codes.

\mysubsection{Electron beams}

In this case, only the electromagnetic interaction plays a role in beam--gas interactions.
The scattering process could be elastic (for example $\mathrm{e}+A\rightarrow \mathrm{e}+A$)
or inelastic ($\mathrm{e}+A\rightarrow \mathrm{e}+X$), see  for example Ref.
\cite{Donnelly}.
        %\begin{equation}%
        % \dyfr{d^2\sigma}{d\Omega\,d\Epr} %(e+Z\rightarrow e^\prime + X)
        %      = \dyfr{(\alpha\hbarc)^2\, \cos^2(\thetahalf)}{M_A\,4E^2\sin^4(\thetahalf)} \,
        %        (W_2(Q^2,\omega) + 2\, \tan^2(\thetahalf) \cdot W_1(Q^2,\omega) )
        % %\dyfr{d\sigma}{d\theta\,d\Epr}(e+Z\rightarrow e^\prime + X)
        % %     = \dyfr{4\pi\, (\alpha\hbarc)^2\, \cos^3(\thetahalf)}{4E^2} \, f(\theta,\Epr)
        %\end{equation}
        %$W_1$  and $W_2$ are structure functions that describe the electromagnetic distribution
        %in the nucleus, and $M_A \approx A\, \mN$.
        %  At small momentum transfers  (small $\theta$, hence only $W_2$ matters)
        %  $W_2 \propto Z^2$   for nuclear only
        %  cross section diverges at very small scattering angles
        % $Z$ = charge of nucleus A in residual gas
Screening of the nuclear charge by the atomic electrons can be important,
see for example  Ref. \cite{Tsai}.
The electron can also scatter directly off the atomic electrons,
a process usually called `M{\o}ller scattering' ($\mathrm{e}^- + \mathrm{e}^- \rightarrow \mathrm{e}^- + \mathrm{e}^-$).
Radiative processes can also be substantial, such as
`Bremsstrahlung' ($\mathrm{e}^- + {\rm Coulomb~field} \rightarrow \mathrm{e}^- + \gamma $),
`pair production'  ($\mathrm{e}^- + {\rm Coulomb~field} \rightarrow \mathrm{e}^- + \mathrm{e}^+ + \mathrm{e}^- $).
In the case of positron beams, `M{\o}ller scattering' becomes `Bhabha scattering'
($\mathrm{e}^+ + \mathrm{e}^- \rightarrow \mathrm{e}^+ + \mathrm{e}^-$) and one also has the process of `annihilation'
($\mathrm{e}^+ + \mathrm{e}^- \rightarrow 2\gamma$).

As an example, we give here formulae for the simplest (and often dominating) process,
namely elastic electron--proton scattering in the Born approximation (no radiative corrections),
as we will also need it later in Section~\ref{Beam--gas imaging}.

\mysubsubsection{Example: elastic electron--proton scattering cross-section}

\begin{figure}
\begin{center}
 \includegraphics[width=0.45\textwidth]{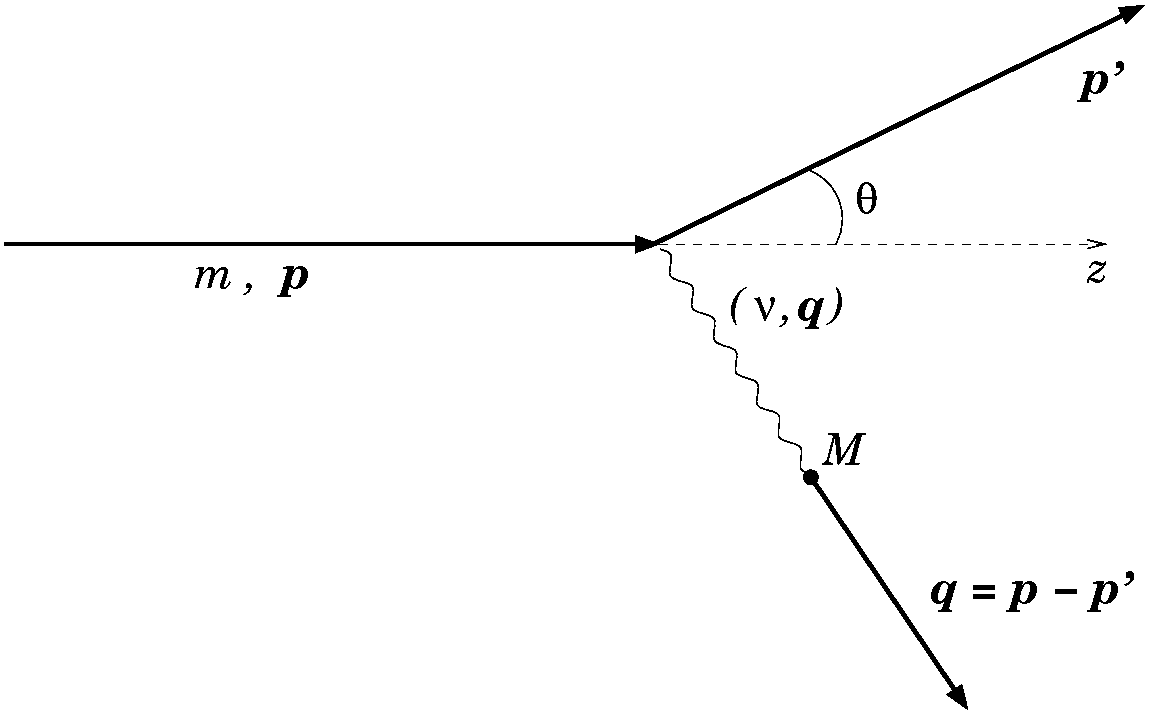}
\caption{Definition of elastic $\mathrm{e}+\mathrm{p}$ scattering in Born approximation}
\label{fig:ep_diagram}
\end{center}
\end{figure}

The process of elastic $\mathrm{e}+ \mathrm{p}$ scattering in the Born approximation is schematically depicted
 in Fig.~\ref{fig:ep_diagram}.

%Note about units: it is customary to give momentum and masses as energies, which is the
%same as setting $c=1$.
%

In the fixed-target frame (FTF) and neglecting the electron mass $m$ ,
the elastic  $\mathrm{e}+\mathrm{p}$ differential cross-section is (see,
for example, Ref. \cite{ep-elastic})
\begin{equation}%
      \frac{\mathrm{d}\sigma}{\mathrm{d}\Omega} =
       \frac{(\alpha \, \hbarc)^2\, \csq}{4E^2\,\ssqsq} \,\frac{E^\prime}{E}~
     % =\frac{( \alpha \, \hbarc)^2\, E^\prime    \, \csq}{4E^3\, \sqq}~
       \frac{\gep^2 +  \tau \, \big(1+2\, (1+\tau)\, \tsq \big)\, \gmp^2}{1+\tau}
       \, ,
\end{equation}
where $\theta$ is the polar electron angle after scattering %in the FTF, relative to the beam axis
and  $\mathrm{d}\Omega =  \mathrm{d}\phi\, \dcos\theta$  is the solid angle for the scattered electron in the FTF.
Integrating over the azimuthal angle $\phi$ gives an extra factor of $2\pi$.
In the Born approximation, a single virtual photon is exchanged between electron and proton.
Its energy is $\nu = E-E^\prime $ with $E$ ($E^\prime$) the electron energies before (after) scattering
in the FTF and its three-momentum transfer is $\mathbf{q} = \mathbf{p} - \mathbf{p^\prime}$ with
$\mathbf{p}$ ($\mathbf{p^\prime}$) the electron momenta before (after) scattering in the FTF.
The scattered electron energy depends on the scattering angle as $E/E^\prime= 1+(2E/\mN c^2)\ssq$.
The four-momentum transfer squared is $q^2=(\nu/c)^2 - \mathbf{q}^2 < 0$ and
one usually defines the positive quantity $Q^2 = -q^2 \approx (4EE^\prime/c^2)\,\ssq$.
The variable $\tau$ is defined as $\tau  = (Q/2\mN c)^2 $. %= (EE^\prime/\mN^2 c^4)\,\ssq$.
The constant $\alpha \approx 1/137 \approx 0.0073$  is the
fine-structure constant (which defines the electromagnetic coupling constant)
and $\hbarc \approx 0.1973~{\rm GeV\,fm}$ is from the Planck constant $h$, $\bar{h} = h/2\pi$.

\begin{figure}
 \includegraphics[width=0.45\textwidth]{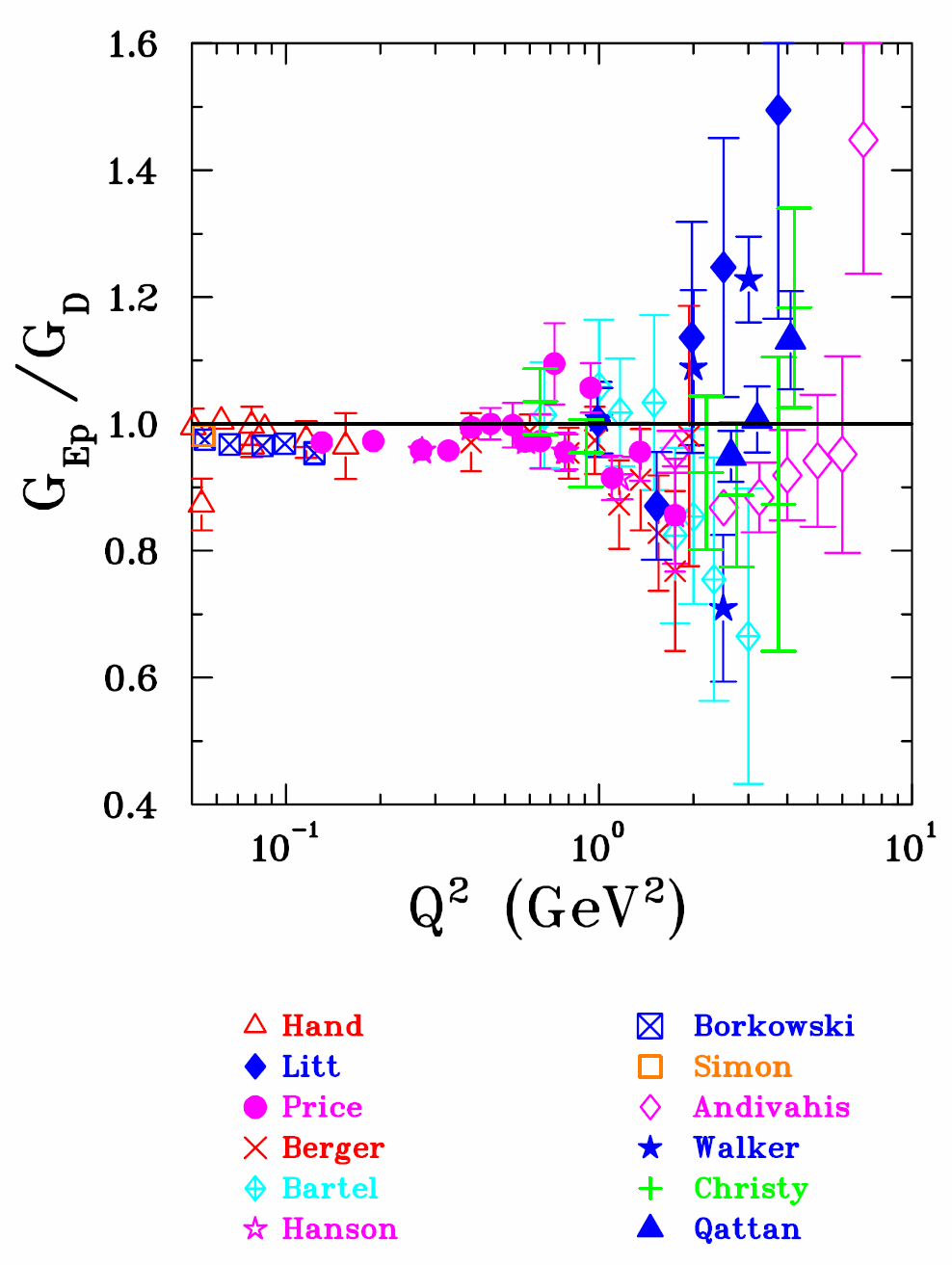}
 \includegraphics[width=0.45\textwidth]{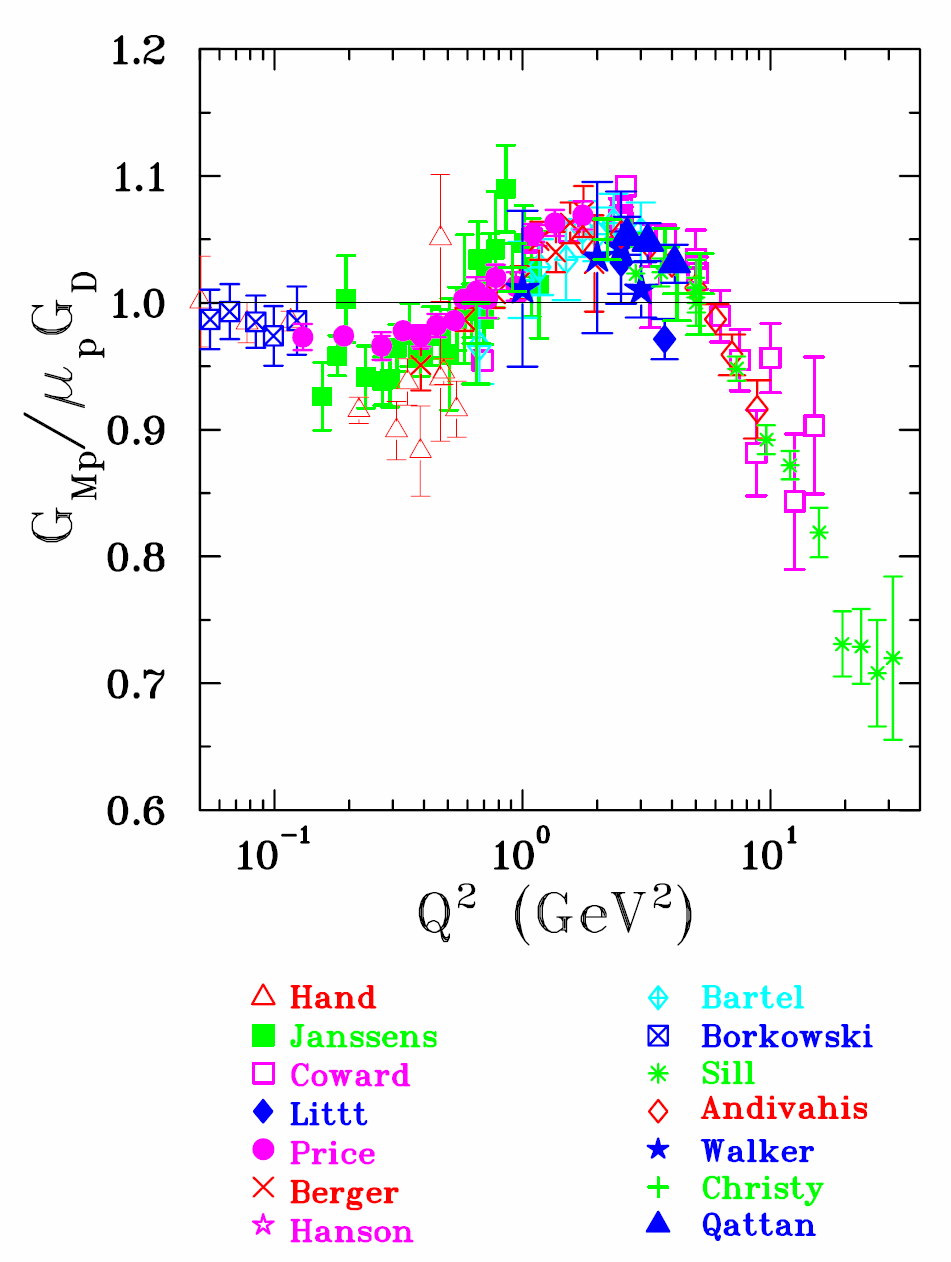}
\caption{Data for $\gep$ and $\gmp$ relative to the dipole formula
  as a function of four-momentum transfer $Q^2$
  obtained by the Rosenbluth cross-section method.
 Figures taken from Scholarpedia~\protect\cite{ep-emff}
  (visit the reference website for a full description).
}\label{fig:ep-emff}
\end{figure}

Depending on its momentum, the virtual photon can resolve the electromagnetic structure of
the proton. This is reflected in
the electric and magnetic proton form factors $\gep(Q^2)$, $\gmp(Q^2)$ which depend only
on the four-momentum transfer squared. They describe the charge and magnetic
distribution in the proton as seen by the virtual photon.
They are measured over a large range of $Q^2$ and, as shown in Fig.~\ref{fig:ep-emff},
are quite accurately rendered by the `dipole formula'~(see, for example,
Ref. \cite{ep-elastic})
\begin{equation}%
  \gep \approx G_D = \Big( 1+\frac{Q^2}{0.71~{\rm GeV}^2/c^2}\Big)^{-2}
  \quad \quad \quad
  \gmp \approx 2.79\, \gep  ~.
\end{equation}
More accurate fits of experimental data can be found in the literature.
In the $Q^2$ range between 0.01 to $0.04~{\rm GeV}^2/c^2$, the form factor $\gep$ varies from 0.97 to 0.9.

For heavier nuclear targets, $ \gep$ and $\gmp$ are replaced by nuclear structure functions.
In the range of small $Q^2$, which usually dominates the cross-section, the longitudinal
structure function prevails and approaches the charge number $Z$. % and magnetic moment $\mu_A$ of the nucleus.
%Because of the kinematic term in front of the , the cross section is entirely dominated
%by the electric form factor.
Thus, for $Q^2\rightarrow 0$, the cross-section approaches
\begin{equation}%
      \frac{\mathrm{d}\sigma}{\mathrm{d}\Omega} =
       \frac{(Z\, \alpha \, \hbarc)^2\, \csq}{4E^2\,\ssqsq} \,\frac{E^\prime}{E}~
       \frac{1}{1+\tau}~.
\end{equation}
One notes that the cross-section diverges for small scattering angles.
Indeed, to obtain a total cross-section (integrated over scattering angles),
the screening of the charge of the nucleus
by the atomic electrons must be taken into account, see \eg Ref.~\cite{Tsai}.

\mysection{Beam--gas losses and beam lifetime}

   Losses due to beam--gas collisions, via some process with cross-section $\sigma_{\rm proc}$,
   in a cyclical accelerator with constant static pressure are characterized by the
   decay rate of the bunch population $N(t)$, which is just the interaction rate $R$
   \begin{equation}
     -\frac{\mathrm{d}N}{\mathrm{d}t} = R = N(t) \cdot \sigma_{\rm proc} \, f\, \cdot \tthick  = \frac{N(t)}{\tau} \, ,
   \end{equation}
   where we defined
   $ \tau^{-1} = \sigma_{\rm proc}\, f\, \tthick$ and $\tthick$  is the gas areal density,
   see Section~\ref{The nature of beam--gas interactions}, that is traversed by the bunch at every turn.
   Assuming that the beam--gas interaction is the only source of bunch population losses,
   the time evolution is
   \begin{equation}%
       N(t) = N(0) \cdot \mathrm{e}^{-t/\tau}
   \end{equation}
   and $\tau$ is the bunch population {\hl lifetime}. % of the bunch population $N(t)$.

   As a numerical example, consider a configuration close to the LHC with proton beams.
   We take a residual pressure $\press = 10^{-9}~$mbar (1~mbar = 100~Pa), composed predominantly of hydrogen
   (H$_2$), at $T = 5$~K, over 27~km. The volume density $\rho$ is obtained from
   the ideal gas law,
   \begin{equation}
      \press = \rho \, k_\mathrm{B}\, T \quad\Rightarrow\quad \rho = \frac{\press}{k_\mathrm{B}\, T} \approx 1.5\times 10^9~{\rm H}_2/{\rm cm}^3~.
   \end{equation}
   This is the concentration of {\hl molecules}. For atoms, one multiplies by two.
   Let us consider the $\mathrm{p} +\mathrm{p}$ inelastic cross-section at 7~TeV, $\sigma_{\rm proc} = 65~$mb,
   see Fig.~\ref{fig:pp-cross-section},
   and assume a revolution frequency $f =~11245~{\rm Hz}$. Then
   \begin{equation}%
    \tau = ( 65 \times10^{-27}~{\rm cm}^2\times 11245~{\rm Hz}\times
               3\times 10^9~{\rm cm}^{-3}\times 2.7\times 10^6~{\rm cm} )^{-1}
         =  1.7\times 10^5 ~{\rm s} = 47 ~{\rm h}~.
   \end{equation}
   Ideally, one would like to lose all beam particles at the experiments, \ie the interaction points,
   and not by beam--gas interactions or other types of losses.
   Consuming the particle bunches by collisions is usually called `burn off'
   and it causes a decay rate for each of the colliding bunch populations $N_1$ and $N_2$
   equal to the colliding-bunch interaction rate $R$
   %\begin{empheq}[box={\mymath[drop lifted shadow, sharp corners]}]{equation}
   \begin{equation}
       -\frac{\mathrm{d}N_1}{\mathrm{d}t} = -\frac{\mathrm{d}N_2}{\mathrm{d}t} = R = \sigma_{\rm burnoff} \, L_{\rm spec}\,  N_1(t)\,  N_2(t) \, ,
         \label{eq:burnoff}
   \end{equation}
   with $\sigma_{\rm burnoff}$ being
   the cross-section of any interaction that causes loss (from the accelerator point of view!)
   of two interacting protons (usually, $\sigma_{\rm burnoff}$ is approximately or a bit less than
    the total cross-section) and
   with  $L_{\rm spec}$ being the `specific luminosity', \ie the luminosity divided by
   the bunch popu\-lation product.
   If one assumes two identical Gaussian bunches, with transverse sizes $\sigma_x$ and $\sigma_y$,
   colliding with zero crossing angle and no transverse offsets, then one has
   $L_{\rm spec} =  f/(4\pi\sigma_x\sigma_y)$ (See, for example, Ref. \cite{HerrMuratori}).
    %\hfill\gray{see lecture \ref{Lec:Introduction}}
   The difference between the populations is constant, since, for each interaction, each bunch
   loses one proton, and indeed one sees that $(\mathrm{d}N_1-\mathrm{d}N_2)/\mathrm{d}t = \mathrm{d}(N_1-N_2)/\mathrm{d}t = 0$.
   If $L_{\rm spec}$ is constant (time-independent), \ie the luminosity decay is only due to
   bunch population losses, while the bunch shapes and overlap do not change with time, then
   Eq.~(\ref{eq:burnoff}) can be easily solved.
   Writing $C \equiv  \sigma_{\rm burnoff} \, L_{\rm spec}$, two cases are considered.
   If the initial population difference is not zero, $\Delta N \equiv N_1(t=0) - N_2(t=0) \neq 0$, then
   for $(i,j)=(1,2)$ or $(2,1)$
 % \begin{equation}
 %     N_2(t) = N_1(t) - \Delta N
 %     = \dyfr{\Delta N}{\dyfr{N_1(0)}{N_2(0)}\,e^{t/\tau_{\rm burnoff}} - 1}
 %     \quad\quad\quad\tau_{\rm burnoff} = (\Delta N \cdot C)^{-1}~,
 % \end{equation}
 % \begin{equation}
 %     N_2(t) %= N_1(t) - \Delta N
 %     = \dyfr{\Delta N\,N_2(0)\,e^{-t/\tau_{{\rm burnoff},1}}}{N_1(0)\,e^{-t/\tau_{{\rm burnoff},2}} -
 %         N_2(0)\,e^{-t/\tau_{{\rm burnoff},1} }}
 %     \quad\quad\quad\tau_{{\rm burnoff},i} = (N_i(0) \cdot C)^{-1}~,
 % \end{equation}
 % \begin{equation}
 %     N_1(t)
 %    =  \dyfr{\Delta N\,N_1(0)\,e^{-t/\tau_{{\rm burnoff},2}}}{N_1(0)\,e^{-t/\tau_{{\rm burnoff},2}} -
 %       N_2(0)\,e^{-t/\tau_{{\rm burnoff},1} }}
 %     \quad\quad\quad\tau_{{\rm burnoff},i} = (N_i(0) \cdot C)^{-1}~,
 % \end{equation}
   \begin{equation}
       N_i(t)
      =  \frac{\Delta N \, N_i(0) \mathrm{e}^{-t/\tau_{{\rm burnoff},j}}}{N_i(0)\,\mathrm{e}^{-t/\tau_{{\rm burnoff},j}} -
      N_j(0)\,\mathrm{e}^{-t/\tau_{{\rm burnoff},i} }}
         \quad\quad\quad\tau_{{\rm burnoff},i} = (N_i(0) \cdot C)^{-1}~,
   \end{equation}
   while if $\Delta N = 0$ and $N_1(t=0) = N_2(t=0) \equiv N_0$,
   then simply
   \begin{equation}
       \frac{\mathrm{d}N}{\mathrm{d}t} = -C\, N^2(t)
       \quad\Rightarrow\quad
       N(t) = \frac{N_0}{1 + t/\tau_{\rm burnoff}}
       \quad\quad\quad\tau_{\rm burnoff} = (N_0 \cdot C)^{-1}~.
   \end{equation}
   %The value $\tau_{\rm burnoff} = (C\, N_0)^{-1}$ is the  half life of $N(t)$.
   Continuing with the same example, we consider the total cross-section $\sigma_{\rm burnoff} = 100~$mb,
   $f = 11245~{\rm Hz}$, as before, $N_0 = 1.2 \times 10^{11}~$protons, $\sigma_x = \sigma_y = 11\Uum$
   and two equally eager experiments, which doubles the interaction rate (as if using $\sigma_{\rm burnoff} = 200~$mb!),
   one can calculate $\tau_{\rm burnoff} = 15~$h, to be compared with the beam--gas lifetime
   $\tau  =  47 ~{\rm h}.$
   Usually, one will design the accelerating machine such that $\tau$ (beam--gas) is considerably
   larger than $\tau_{\rm burnoff}$.

\mysection{Detector background}

In some cases, beam--gas interactions in the neighbourhood of an experiment
(or any particle detection system) are
  the source of undesired background rates in the detectors.
Detailed background studies have been performed at the LHC, see for example Refs.~\protect\cite{ATLAS-BIB,ATLAS-BIB2,Gibson}.
Here, we focus on a notable example at the LHC, given by the ALICE detector~\cite{ALICE}.
ALICE was designed for relatively low luminosity compared with  the other large LHC experiments
(ATLAS, CMS, and LHCb).
  ALICE has two main p + p running modes, corresponding to different trigger configurations
   (and physics goals):
   \begin{itemize}
     \item `Minimum bias' acquisition: $2\times 10^{29}\Ucm^{-2}\!\Us^{-1}$, rate $\approx 150$~kHz,
     \item `Rare events' acquisition: $8\times 10^{30}\Ucm^{-2}\!\Us^{-1}$, rate $\approx 600$~kHz.
   \end{itemize}

   ATLAS and CMS operate at a luminosity around $10^{34}\Ucm^{-2}\!\Us^{-1}$.
  In p + p collider mode, the LHC runs primarily for the ATLAS, CMS, and LHCb experiments,
  but ALICE requires to take data as well.
  The factor 10,000 mismatch in luminosity requirement engenders serious challenges
  for the LHC machine operation.
  The maximum bunch intensity and number of bunches must be utilized, which
  causes high beam--gas interactions rates, owing to the residual gas.
  ATLAS, CMS, and LHCb are designed to cope with high p + p interaction rates,
  and therefore are rather immune to the added beam--gas interaction rates naturally
  occurring in the vicinity of the experiments, provided these rates remain well below
   the p + p interaction rate. ALICE is different.

  The driving physics running mode of ALICE is with lead ion beams, for which
  the design lumi\-nosity is many orders of magnitude smaller than for p +
  p (although
   the particle multiplicity per inter\-action is much larger in the lead--lead collisions!).
   The main tracking detector in ALICE is a time projection chamber (TPC)~\cite{ALICE-TPC}.
  It is essentially a gas volume
  composed of two concentric cylinders with an inner (outer) diameter of 1.6 (5)~m, split at
  mid-length by an equipotential plane and terminated at each end by a detector plane.
  The length is $2.5~$m on each side of the mid-plane.
  The gas volume is exposed to a high voltage of around 100~kV, generating an electric field
  of about 40~kV/m along the cylinder axis.
  In such a detector, electrons generated by ionization of gas through the passage of a high-energy
  charged particle will drift along the electric field lines until they reach the actual detection
  device situated at the extremities of the gas volume.
  In the ALICE case, the drift time can be as long as $90\Uus$, which is the time for about
  one LHC bunch revolution.
  The high voltage in the field cage is
  protected with current trip limits of about $7\UuA$, which corresponds to a particle
  rate of  about 500~kHz. This rate is reached at a p + p luminosity of around $7\times 10^{30}\Ucm^{-2}\!\Us^{-1}$.
   Two main issues were encountered with the ALICE detector in the initial run of the LHC:
   (a) the minimum bias trigger accepted beam--gas events, which resulted in excessive rates
,   and (b) the beam--gas rates were so high that they precluded turning on the high voltage of the TPC.

   As an exercise, let us estimate (roughly) what would be the ALICE pressure requirement for the
  insertion region around the interaction point.
  Assume beam--gas interactions originating from up to $\Delta z~= 100$~m away leave tracks in the TPC
  (and, eventually, could induce a minimum bias trigger).
  Suppose that the residual pressure profile is flat and mainly due to hydrogen molecules
  at room temperature $T= 293\UK$.
  Further assume nominal LHC conditions ($N = 1.1\times 10^{11}$\,p/bunch, $n_\mathrm{b} = 2808$ bunches at 7~TeV).
  Starting from the rate $R = \sigma_{{\rm inel},\mathrm{p}+\mathrm{p}}\,n_\mathrm{b}\, N \, \, f\, \tthick$,
        see Eqs.~\ref{eq:rate1} and \ref{eq:rate2},
  one can deduce that, in order to  have a contribution of beam--gas events to the triggers in ALICE of
  less than 50~kHz, from $R < 50~$kHz and $P({\rm H_2}) = \frac{1}{2}\, k_\mathrm{B}\, T \, \tthick/\Delta z$ one deduces that
  the hydrogen pressure $P({\rm H_2})$ should obey
  \begin{equation}
      P({\rm H_2}) %= \frac{1}{2}\, k_B\, T \, \dyfr{\tthick}{\Delta z}
      < \frac{ \frac{1}{2} \times 1.38\times 10^{23}~\frac{\rm J}{\rm K}\times 293~{\rm K}\times
      50~{\rm kHz}}{100~{\rm m}\times3\times 10^{14}\times 11245~{\rm Hz}\times 4.5\times 10^{-30}~{\rm m}^2}
      = 5\times 10^{-10}~{\rm mbar} \, ,
      %=5\cdot 10^{-8}~{\rm Pa}= 6.7\cdot 10^{-10}~{\rm mbar}$
   \end{equation}
    % \small 1~mbar = 100~Pa
   where  the value $\sigma_{{\rm inel}, \mathrm{p}+ \mathrm{p}} = 45$~mb  has been used.
  %= 55e-27~{\rm cm}^2\, 3e14 \, 11245~{\rm Hz}\, \rho\, \Delta z < 50~kHz$

\Ignore{
   At the LHC, several mitigating measures were implemented to reduce the efefct of beam-gas background on the
  ALICE experiment:
    \begin{itemize}
    \item proper low SEY\footnote{Secondary Electron Yield, see~\protect\cite{Lec:ThinFilm}.} coatings were applied on
            warm surfaces of critical vacuum chambers,
    \item pumping speed (ion pumps and getter pumps) were added,
    \item solenoids were implemented to reduce electron multipacting,
    \item beam-viewing surfaces were conditioned (scrubbed) with beam,
    \item bunch patterns were optimized to reduce beam-induced vacuum degradation.
    \end{itemize}
    see lecture \cite{Lec:Dynvac}.
    With the result that the pressure was reduced by more than one order of magnitude.
}

\begin{figure}
   \centering
             \includegraphics[width=0.44\textwidth]{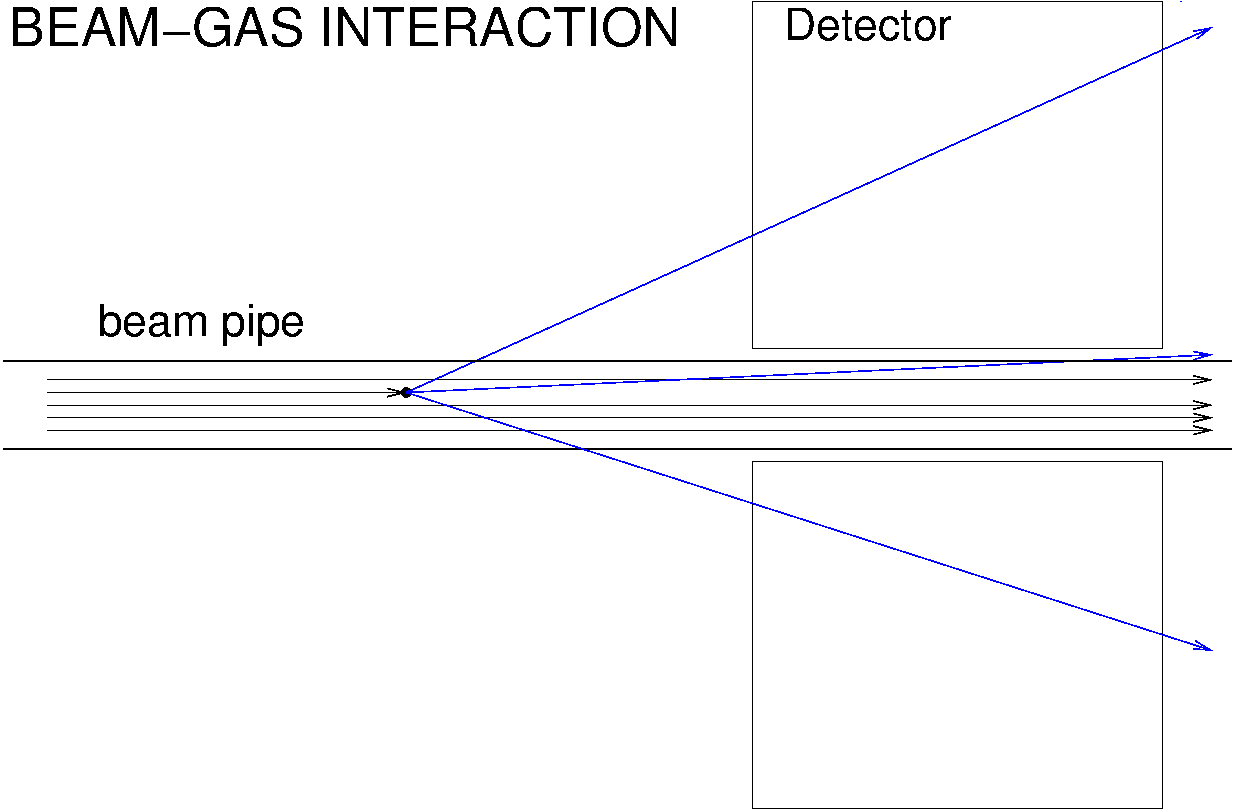}~ ~ ~ ~ ~
             \includegraphics[width=0.44\textwidth]{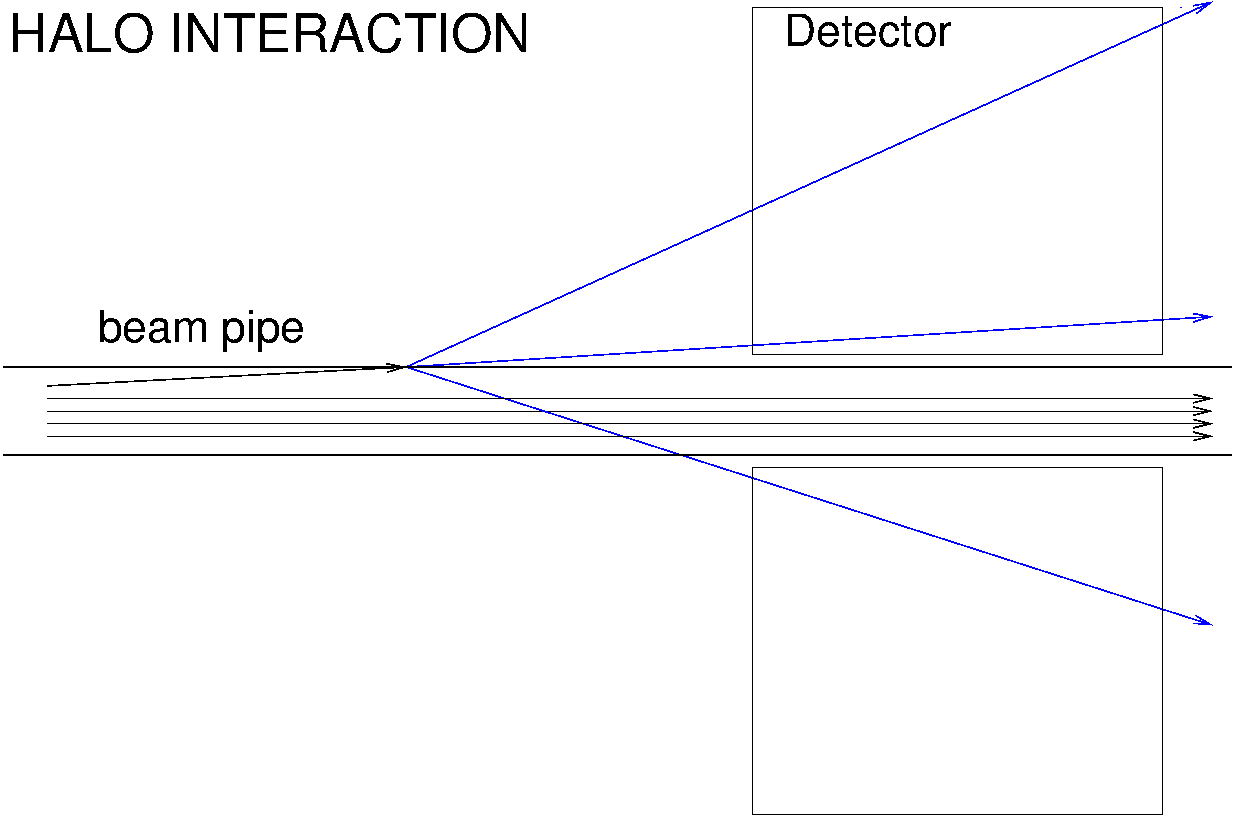}
   \caption{Left, beam--gas interaction; right,  halo particle interacting
   with surrounding material }\label{fig:sketch-beamgas-halo}
   \end{figure}

   Beam--gas interactions can sometimes be confused with halo particle interactions, as shown in Fig.~\ref{fig:sketch-beamgas-halo}.
   A detailed study of the signatures of the background events was carried out to demonstrate that a large fraction originated from beam--gas events
   and to identify the  regions around the experiments that were the sources of these
   events~\cite{ALICE}.

  %\begin{figure}[htb]
  %\begin{center}
  %\includegraphics[width=0.7\textwidth]{figs/ALice_v0av0c.png}
  %\includegraphics[width=0.8\textwidth]{figs/figures_Fig6_2.png}
  % \caption{Top: Sketch explaining the signature of beam-gas and bunch-bunch interactions.
  %          Bottom: weighted average time of flight of the particles detected in ALICE V0-C versus V0-A.
  %                  The time offsets are defined such that bunches cross the IP at 0;
  %                  figure taken from reference~\protect\cite{AliceV0}.}\label{fig:v0}
  %\end{center}
  %\end{figure}

%  \begin{figure}[htb]
%  \begin{center}
%  \includegraphics[width=0.49\textwidth]{figs/182454minbias_pixclus_charge_col_notselbbg.png}
%  \includegraphics[width=0.49\textwidth]{figs/182454background_pixclus_charge_col_selbbg.png}
%   \caption{ATLAS pixel cluster deposited charge versus cluster width (in pseudorapidity space),
%            for Pixel barrel clusters. Left: for data taken with colliding bunches.
%            Right: for data taken with non-colliding bunches.
%            Figures taken from reference~\protect\cite{ATLAS-BIB}.
%            }\label{fig:ATLASclusters}
%  \end{center}
%  \end{figure}

\begin{figure}
   \centering
     \includegraphics[width=0.7\textwidth]{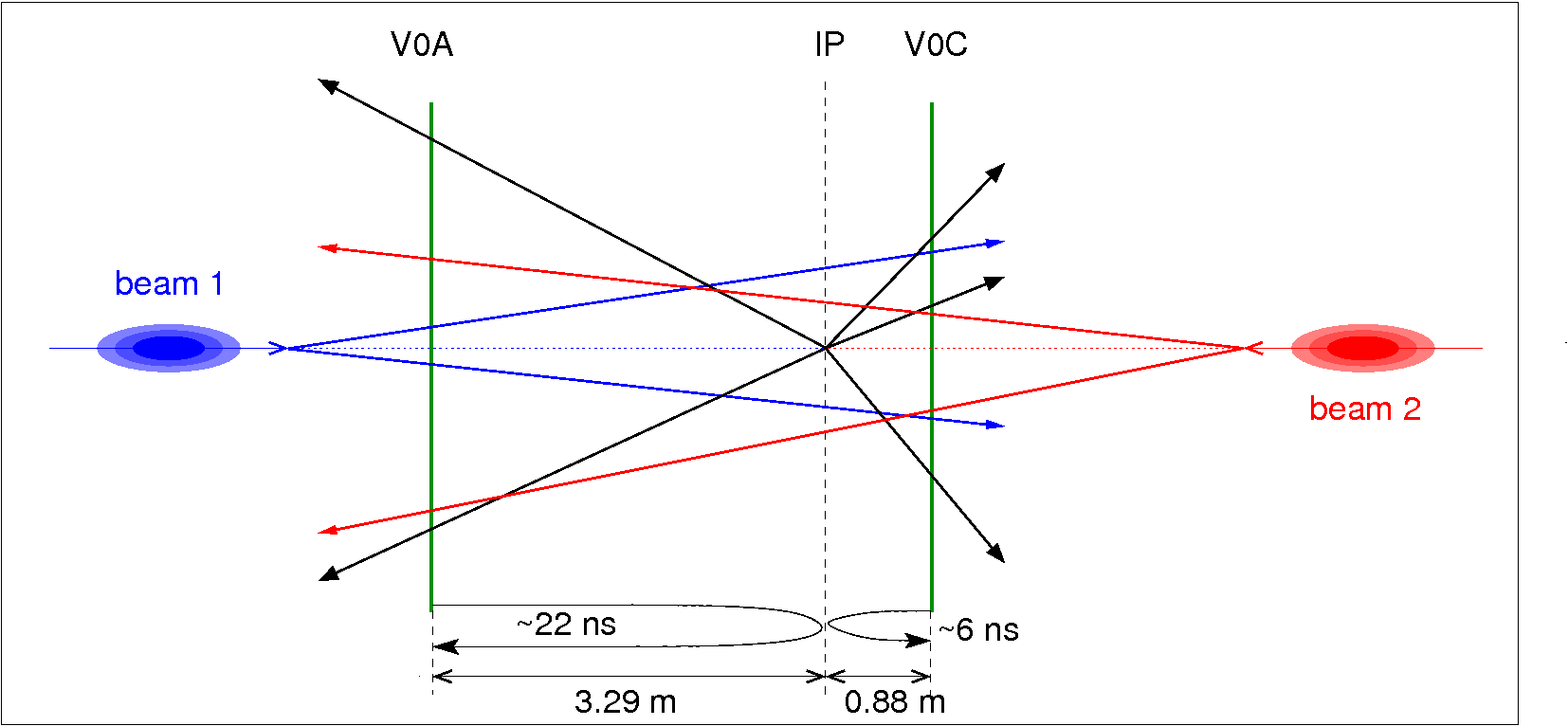}
     \caption{Sketch explaining the arrival times of beam--gas and beam--beam interactions
         relative to the passage of bunches at the IP.}\label{fig:v0-sketch}
   \end{figure}

\begin{figure}
   \centering
   \includegraphics[width=0.8\textwidth]{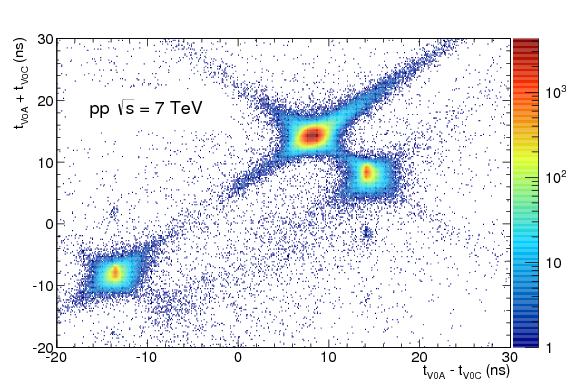} %V0_4-8378.png}
   \caption{Correlation between the sum and difference of signal times in ALICE V0A and V0C detectors.
    Three classes
    of events---beam--beam interactions at (8.3 ns, 14.3 ns), background from beam 1 at ($-$14.3 ns, $-$8.3 ns),
    and background from beam 2 at (14.3 ns, 8.3 ns)---can be clearly distinguished.
Figure taken from Ref.~\protect\cite{ALICE}.}\label{fig:v0-data}
   \end{figure}

   The main signature was the measured time correlation between the V0-A and V0-C detectors~\cite{AliceV0},
   which are scintillator arrays located longitudinally at $-$3.29~m and $+$0.88~m,
 respectively, from the interaction point (IP),
   as depicted
   in the Fig.~\ref{fig:v0-sketch}. Each segment of the detector measures a hit time relative to the
   time at which LHC bunches cross the nominal IP and a weighted average time is formed for the two arrays,
   $t_\mathrm{V0A}$ and $t_\mathrm{V0C}$.
   Figure~\ref{fig:v0-data} shows
   %the weighted average time of flight of the particles detected in the ALICE V0-C and V0-A scintillator arrays.
   the sum $t_\mathrm{V0A}+t_\mathrm{V0C}$ plotted with respect to the time difference $t_\mathrm{V0A}-t_\mathrm{V0C}$.
   The beam1-induced and beam2-induced background interactions %(``p-gas interactions'')
   coming from the left and right sides, respectively, of the sketch  are
   %clearly visible as a blob at the expected timing locations around (-11, +3)~ns
   %and (+11, -3)~ns. The beam-beam interactions (``pp collisions'') are located around (+11, +3)~ns.
   clearly visible as blobs at the expected timing locations around ($-$14.3 ns, $-$8.3 ns)
   and (14.3 ns, 8.3 ns). The beam--beam interactions are located around (8.3~ns, 14.3~ns).

\begin{figure}
   \centering
   \includegraphics[width=0.8\textwidth]{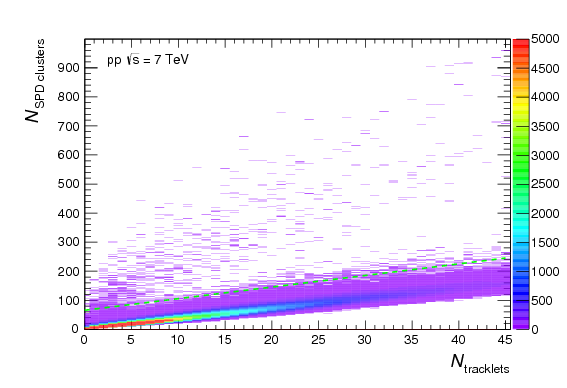}%  spd_clus_104892-8381.png}
    \caption{ %(a) Event display showing two tracks coming at a large polar angle from the IP
              %    into the ALICE silicon pixel detector (SPD) with sensor tiles oriented parallel to the beam axis.
             %(b)
           %  Plot of the ALICE silicon pixel detector (SPD) cluster multiplicity versus the number of
           %  tracklets in the event (courtesy of Antonello Di Mauro).}\label{fig:SPD}
            Correlation between reconstructed ALICE SPD clusters and tracklets. Two bands
            corresponding to the collisions and machine-induced background are visible.
             The dashed cyan line represents the cut used in the off\-line selection:
             events lying in the region above the line are tagged as beam--gas and rejected.
               Figure taken from Ref.~\protect\cite{ALICE}.
          }\label{fig:SPD}
   \end{figure}

   A second signature was the correlation between the number of clusters and tracklets in the ALICE silicon pixel detector (SPD).
   The SPD is composed of thin silicon detector tiles oriented parallel to the beam axis. Therefore, when a track
   originates from a far distance from the IP, it will impinge on the sensor tiles at a shallow angle,
   almost parallel to the beam axis, leaving a large number of hits in the sensor.
   Tracks originating from the IP will typically leave one cluster per sensor.
   This signature is visible in the graph of Fig. \ref{fig:SPD} (right), which shows the
   SPD cluster multiplicity versus the number of tracklets in the event.
   The branch at a larger slope than the main band is due to the distant interactions
   (in this case, beam--gas interactions).
   This signature allowed quantifying and monitoring of the rate due to beam--gas processes.

   Other tricks were used to demonstrate the effect of beam--gas interactions.
   Plotting background rate as a function of the product of  beam intensity and measured residual pressure
   gives another hint. Indeed, if background originated, for example, from halo scattering
   from the beam pipe, the rates could scale with beam intensity, as for beam--gas rates,
    but they would probably not scale with the residual pressure.
   Using the `forwardness' of tracks may also help in disentangling beam--gas processes from,
   for example, longitudinally displaced bunch--bunch interactions.
   If both forward- and backward-moving tracks emerge from a vertex, it is unlikely to be
    a beam--gas interaction.
   Finally, the use of vertexing to identify beam--gas events can be effective.
   If all tracks point to a vertex inside the beam pipe, it is likely to be a beam--gas event,
   as opposed to, for example, a halo scattering event.

     This example is representative of a high-energy proton machine.
     In electron (positron) machines, the sources of background are quite different.
     The beam--gas background is then dominated, usually, by radiative effects  and
     electromagnetic showering. % (see for example~\cite{KEK-BEAST}).

\mysection{Beam--gas imaging}

  Beam--gas interactions are not only a nuisance. They may also be used as a tool
  for imaging the beam properties.
  In the LHCb experiment~\cite{LHCb}, beam--gas interactions are used for many purposes, for instance
  to perform transverse beam profile  measurements, to quantify the ghost charge,
  to cross-check the relative bunch population measurements, to determine the
  absolute luminosity with high precision~\cite{LHCb-BGI,LHCb-lumi} and, as we will develop in the next section,
 to carry out fixed-target physics experiments.

  %   \begin{figure}
  %      \includegraphics[width=0.45\textwidth]{figs/y-LHCb-reoptimized-with-beams_trimmed.png}
  %      \includegraphics[width=0.45\textwidth]{figs/velo_layout_schematic.pdf}
  %      \caption{LHCb, A key detector: {\hl VE}rtex {\hl LO}cator.}
  %   \end{figure}

\begin{figure}
      \centering
          \includegraphics[width=0.48\textwidth]{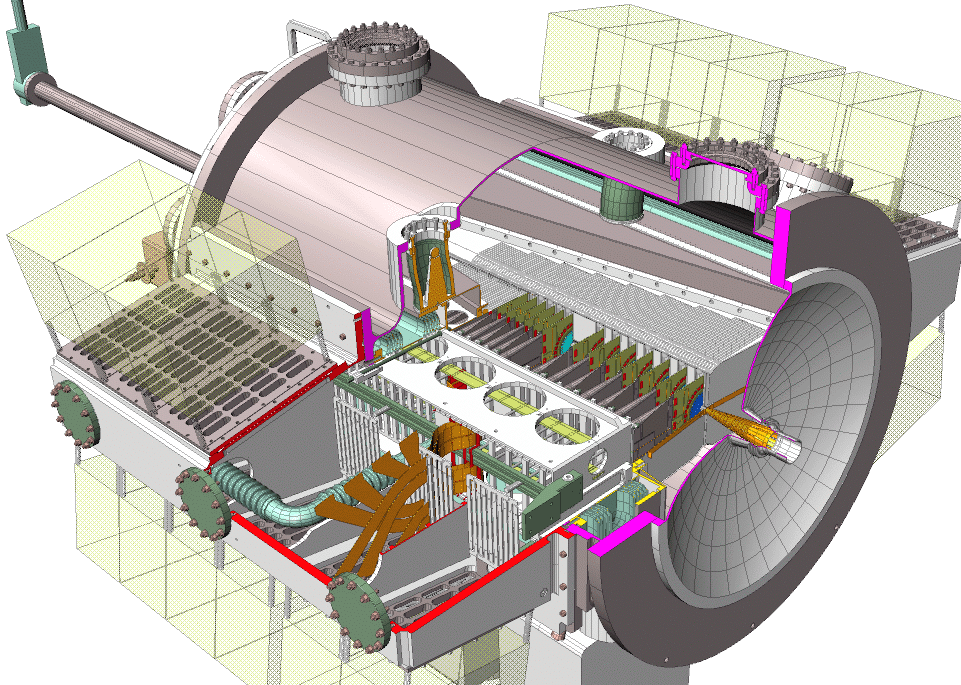}
          \includegraphics[width=0.48\textwidth]{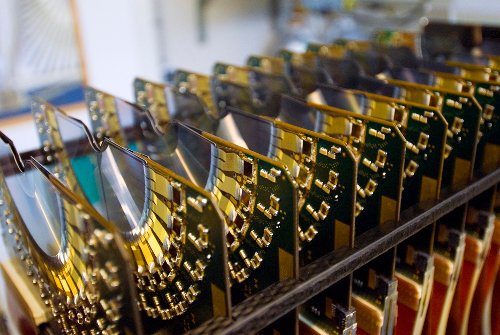}\\
          \includegraphics[width=0.96\textwidth]{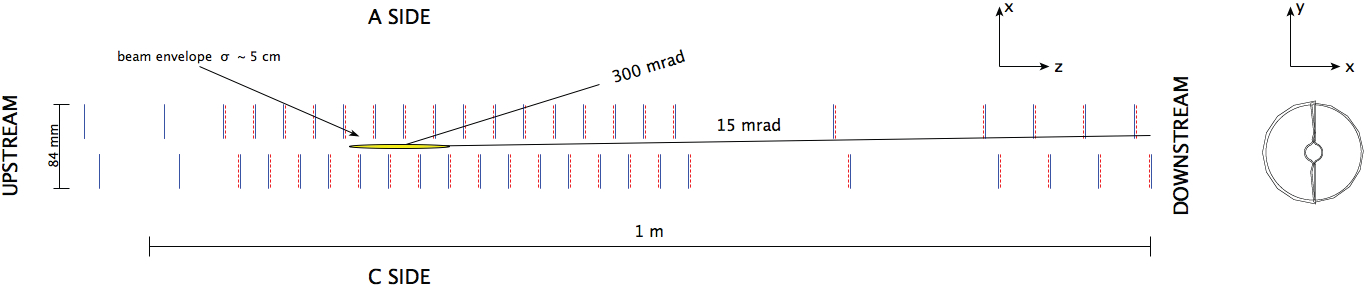}\\
          \caption{Top left: The LHCb VELO vacuum tank. The cut-away view allows the VELO sensors, hybrids,
          and module support on the left-hand side to be seen. Top right: A photograph of one side of the VELO
          during assembly showing the silicon sensors and readout hybrids. Bottom: Cross-section in the $xz$ plane at
          $y = 0$ of the sensors and a view of the sensors in the $xy$ plane.
          The detector is shown in its closed position.
          $R$ ($\phi$) sensors are shown with solid blue (dashed red) lines.
          The modules at positive (negative) $x$ are known as the left or A-side (right or C-side).
          Figures taken from Ref.~\protect\cite{VELO}.}\label{fig:velo}
      \end{figure}

\begin{figure}
      \centering
        \includegraphics[width=0.55\textwidth]{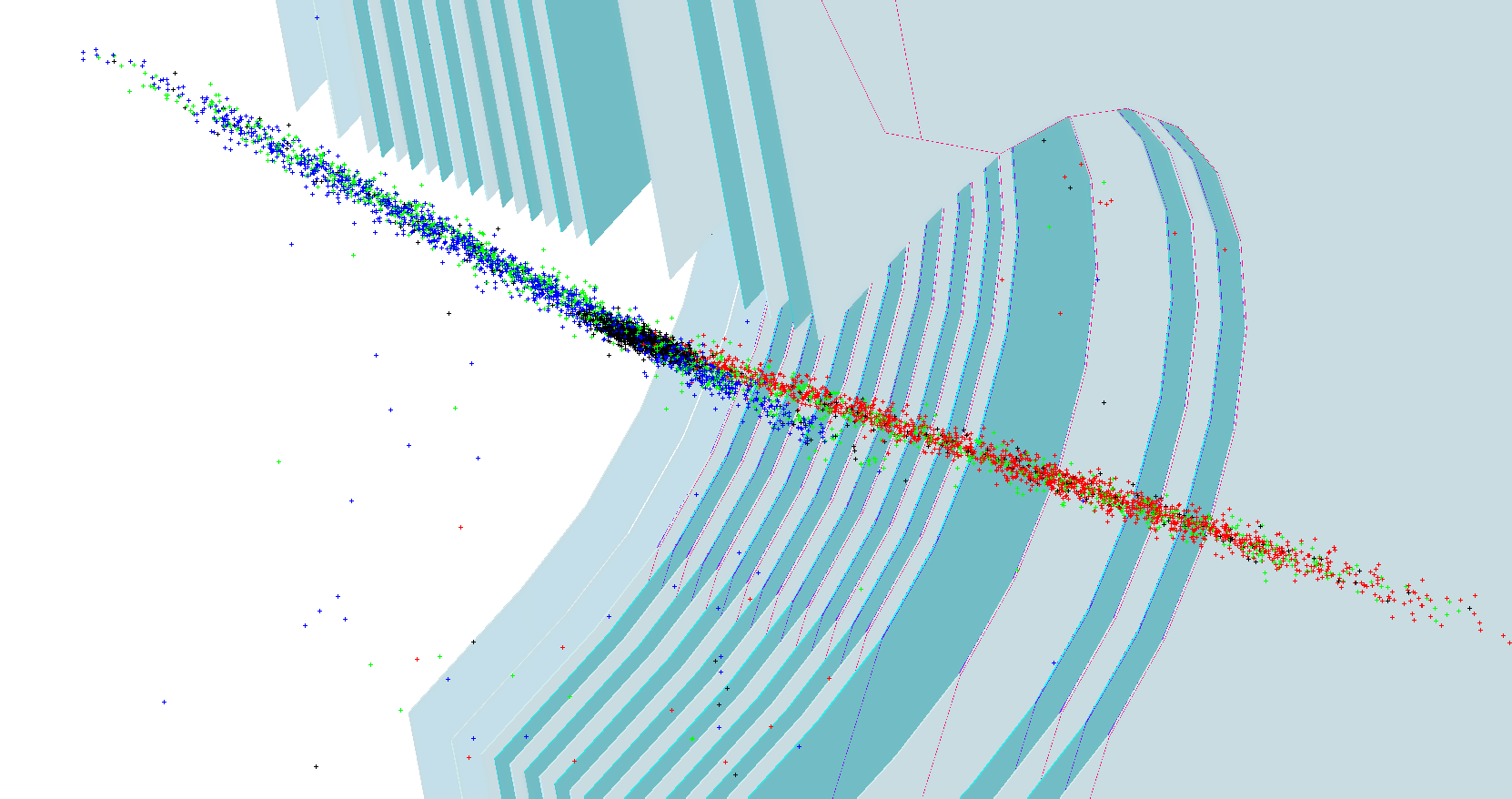}
      \includegraphics[width=0.44\textwidth]{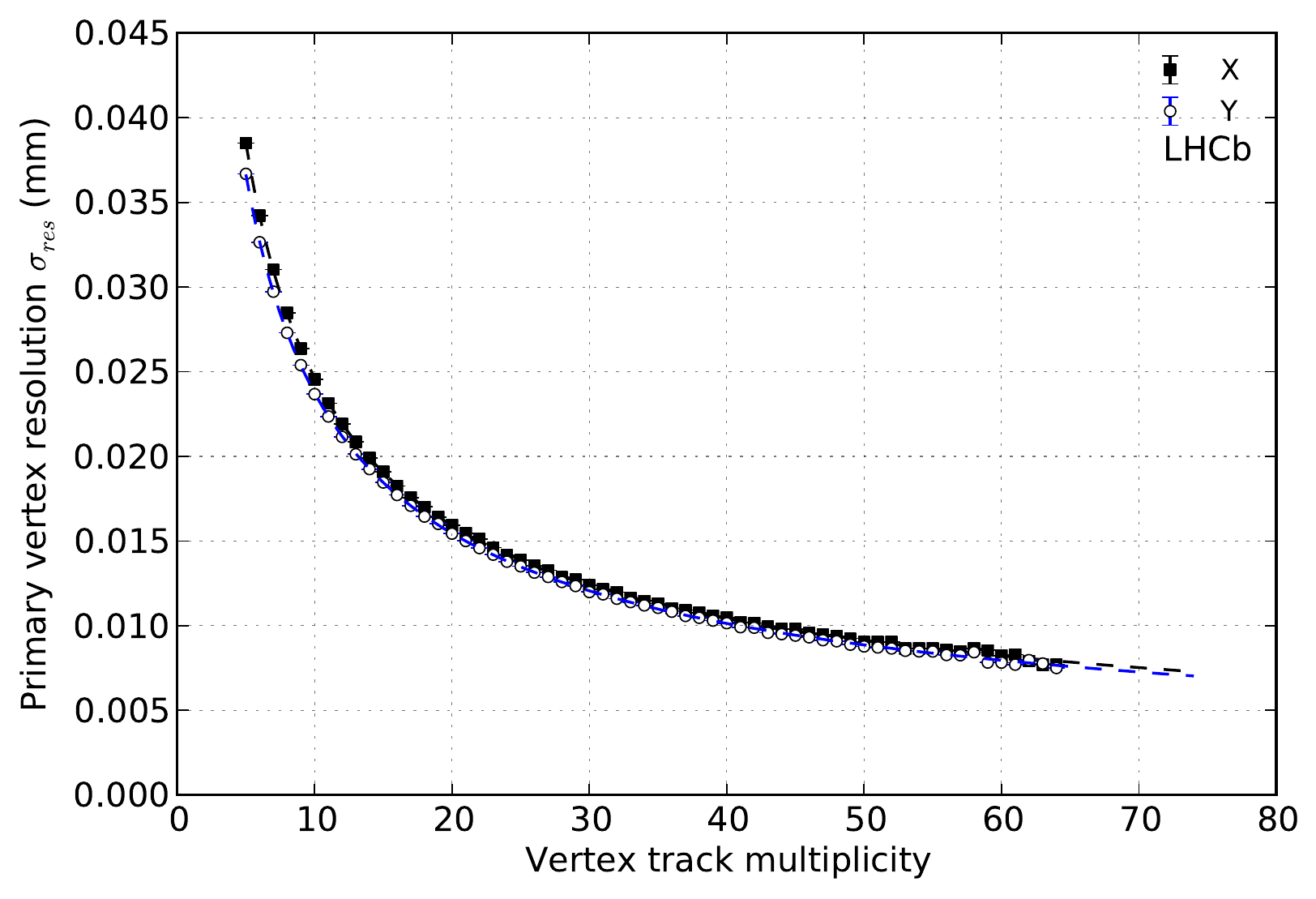}
        \caption{Left: zoomed display of the reconstructed vertices near the IP.
                  Blue, red, green, and black dots represent vertices measured in
              \be, \eb, \bb, {\ee} LHC bunch crossings.
              Right: LHCb VELO {\bb} vertex resolution as a function of vertex track multiplicity.
          Figures taken from Ref.~\protect\cite{LHCb-BGI}.}\label{fig:bgi}
    \end{figure}

  A key detector in all these tasks is the LHCb Vertex Locator (VELO)~\cite{VELO}, see Fig.~\ref{fig:velo},
   which is a precision tracking detector, located in the vicinity of the beams in a secondary vacuum.
   It is composed of vertically oriented silicon microstrip sensors, each
with their active edge at about 8~mm
   from the beams.
   Figure~\ref{fig:bgi} (left) displays reconstructed interaction  vertices from  \be, \eb, \bb, {\ee} LHC bunch
   crossing types. Here, the two letters designate the nominal state of the crossing bunch slots of beam 1 and beam 2.
    A {\tt b} stands for a nominally filled bunch slot and an {\tt e} stands for a nominally empty bunch slot.
   The traces of the two beams and of the luminous region are clearly visible, including the crossing angle.

    The VELO vertex resolution can be seen in Fig.~\ref{fig:bgi} (right) for p + p interactions as a function of
   vertex track multiplicity.
   In addition, the VELO has a good acceptance for beam--gas events.
   %Explain rapidity boost, forward-boosted beam-gas interactions!
   This feature was used to reconstructed the transverse beam profiles, offsets, and angles, and
   combined with the luminous region vertex distribution, to calculate the beam overlap integral
   \begin{equation}
       {\ovlap}  = 2c\,\cossqa\, \int \rho_{1}(x,y,z,t)\,\rho_{2}(x,y,z,t)\, \mathrm{d} x\,\mathrm{d}y\,\mathrm{d}z\, \,\mathrm{d}t
  \end{equation}
  of two counter-rotating bunches (1 and 2) with time- and position-dependent
  density functions $\rho_{1}(x,y,z,t)$ and $\rho_{2}(x,y,z,t)$ that drives the luminosity
   \begin{equation}
         L %= f \,N_1\,N_2\,\, 2c\,\cossqa\, \int \rho_{1}(x,y,z,t)\,\rho_{2}(x,y,z,t)\, dx\,dy\,dz\, \,dt \\[4mm]
           = f \,N_1\,N_2\, {\ovlap}~,
     \label{eq:lumi}
  \end{equation}
  where $N_1$ and $N_2$ are the total number of protons in the bunches and
  $\hca$ is half the crossing angle.

\begin{figure}
      \centering
    \includegraphics[width=0.48\textwidth]{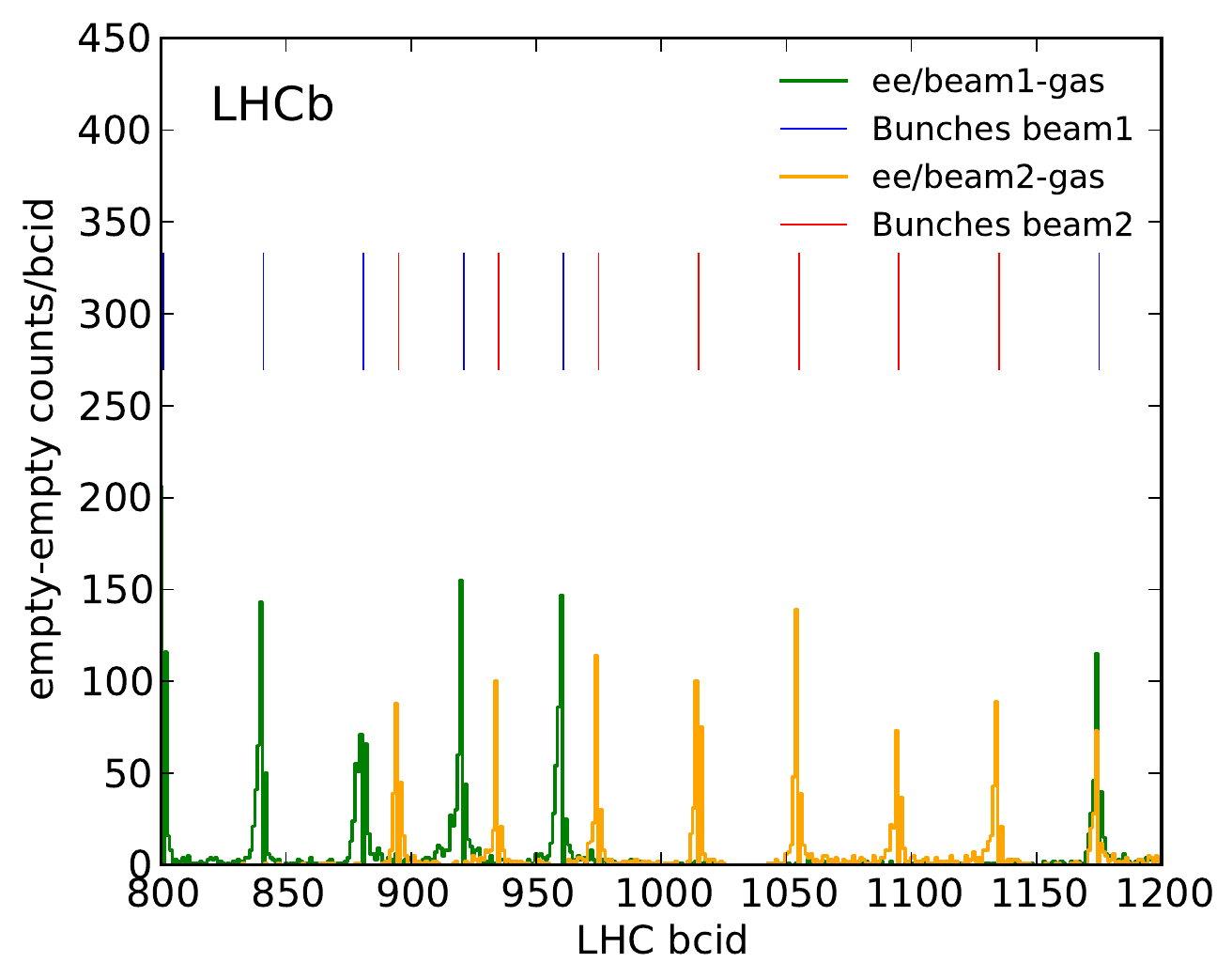}
      \includegraphics[width=0.48\textwidth]{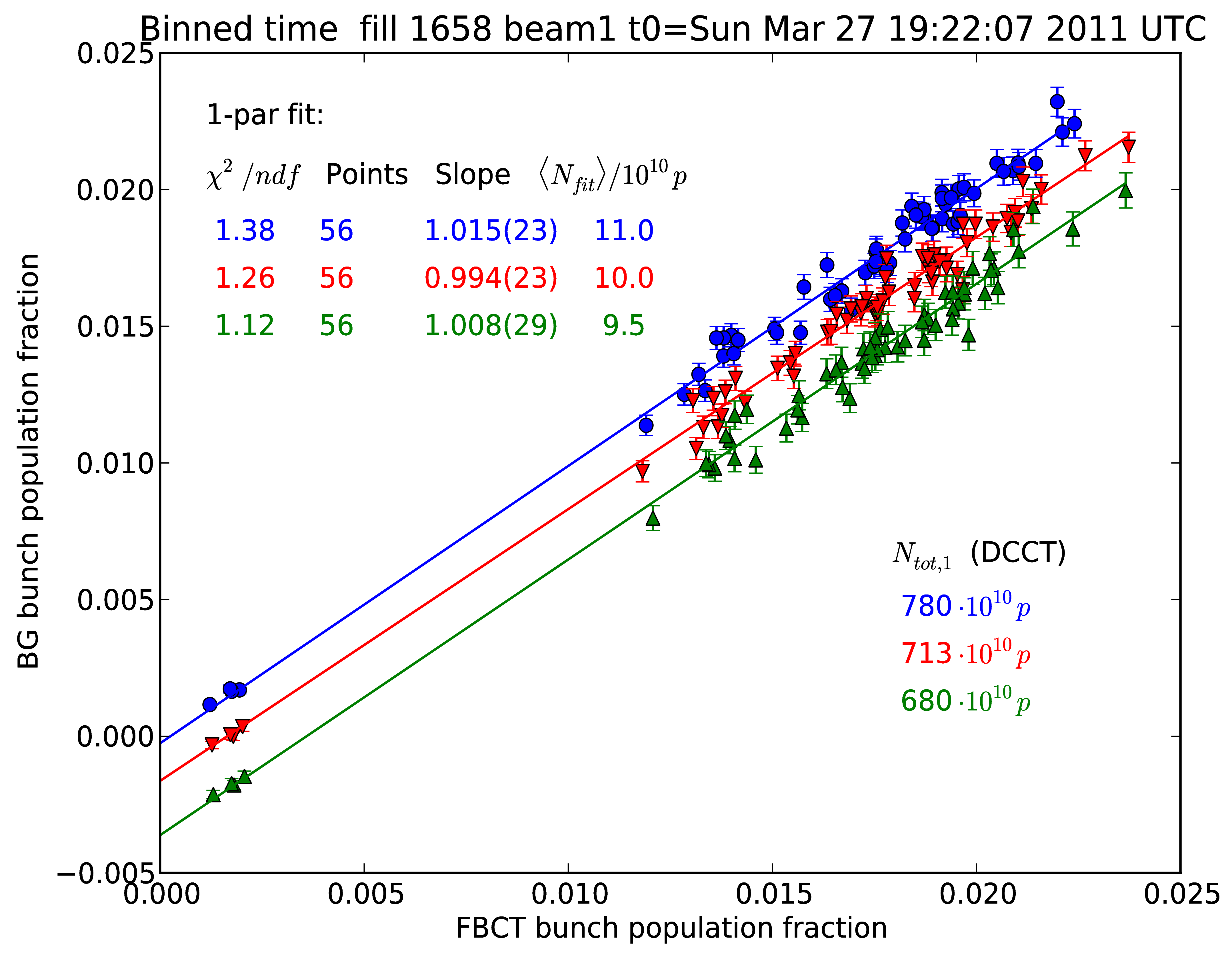}
      \caption{
        Left:
          Example of LHCb  ghost charge measurements by beam--gas rates in nominally empty bunch slots,          integrated over 4 min, shown as a function of the LHC bunch slot number (bcid), see text.
          Taken from Ref.~\protect\cite{LHCb-BGI}.
          Right:
          LHC bunch populations measured by the LHCb beam--gas rates versus
           those measured by the LHC FBCT, averaged over 3 h bins, one colour per time bin
          (for clarity, the data have been shifted down by a fixed amount after each time bin).
         Taken from Ref.~\protect\cite{BCNWG3}.}\label{fig:bgi-fbct}
       \end{figure}

   The normalization of the bunch populations $N_1$ and $N_2$ is crucial for the luminosity determination
   and was obtained from  dedicated LHC devices.
  The total circulating charge was measured using a dir\-ect current--current transformer, which measures precisely the total beam population~\cite{DCCT}.
   A fast bunch current  transformer (FBCT) is used to measure the relative bunch populations~\cite{BCNWG3}.
   However, the FBCT has an intrinsic population threshold below which no population is measured.
   Given the large number of nominally empty bunch slots, even a small unmeasured population
   per slot may result in a substantial error in the bunch population normalization.
   For this reason, the LHCb beam--gas technique was used to count the beam--gas events in every
   bunch slot (filled or empty) and, by comparing the rates, to determine the amount of
   `ghost charge' contained in the sub-threshold slots~\cite{LHCb-BGI}.
   An example result is shown in Fig.~\ref{fig:bgi-fbct}  (left), which shows the beam--gas event
   counts for a 4 min integration time  as a function of LHC bunch slot number (bunch crossing ID).
   The vertical blue and red lines indicate the nominally filled slots
   (slot 1175 is a {\bb} crossing type, the blue line hides the red line).
   The beam--gas rates for these slots are suppressed from the graph
   (they contain several tens of thousands counts).
  Shown are (in green and orange) the beam1- and beam2-gas rates for nominally empty slots.
  Only a zoomed range of slots is shown (from 800 to 1200, out of 3564 LHC slots).
  Here, the ghost charge clustered around the nominally filled bunch slots.
  %\begin{figure}
  %    \includegraphics[width=0.99\textwidth]{figs/LDM_Adam.pdf}
  %    \caption{(courtesy of J. Adam)}
  %\end{figure}
 %In order to normalize the $N_1$ and $N_2$ one needs to know how much of the total population
 %circulates outside the nominal RF buckets.

  The beam--gas rates of nominally filled slots were also used to cross-check the linearity of the FBCTs~\cite{BCNWG3}.
  Examples of LHCb relative bunch populations measurements by beam--gas rates
  are shown in Fig.~\ref{fig:bgi-fbct}.
  Different colours or markers are just different time periods (with an artificial offset for clarity,
   except for the blue).

    To increase the beam--gas interaction rate, and thus the statistics,
    a system was installed, the System for Measuring the Overlap with Gas (SMOG),
    which allows
    one to inject a tiny amount of gas (Ne, He, or Ar) in the VELO beam vacuum
    and thereby increase the pressure from $10^{-9}~$ to $10^{-7}~$mbar.
    Only light noble gases are used because of the  presence of non-evaporable getter  coatings in the vicinity
      of the IP, see Ref. \cite{Lec:Getters}.

\begin{figure}
       \includegraphics[width=0.48\textwidth]{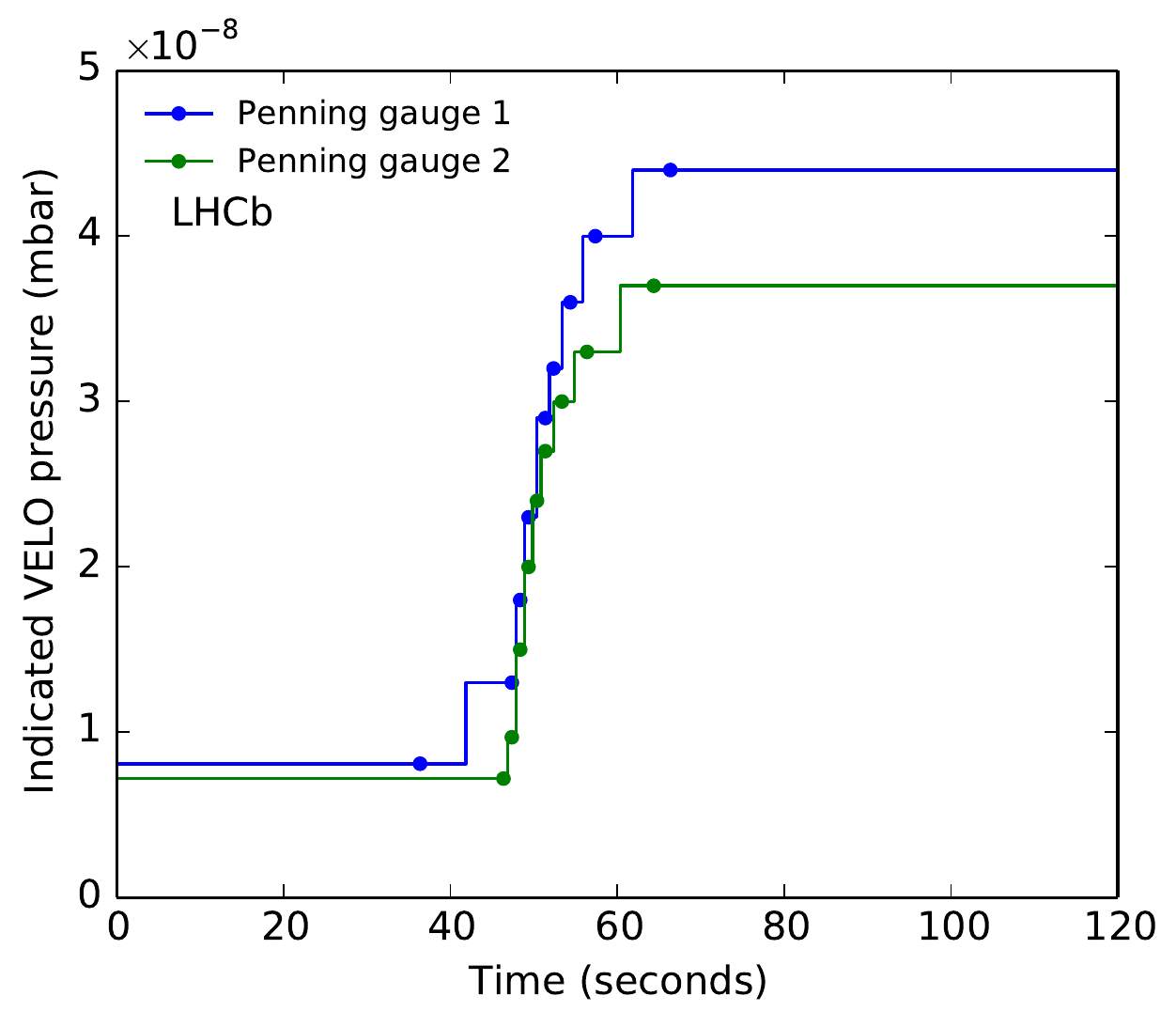}
       \includegraphics[width=0.48\textwidth]{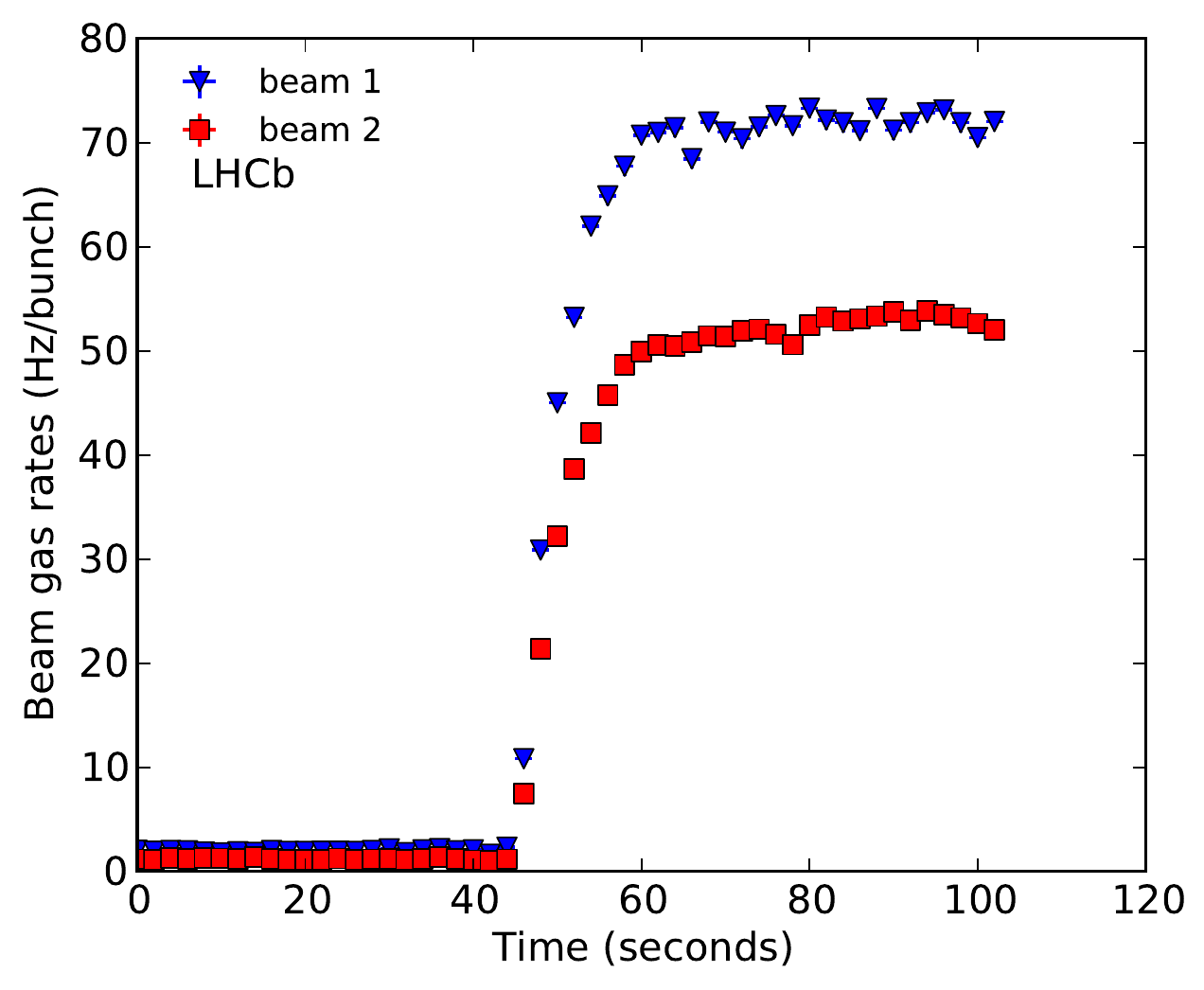}

       \caption{Example of VELO pressure (left) and beam--gas rate (right) as a function of time
                measured at the LHCb experiment.
                Taken from Ref.~\protect\cite{LHCb-BGI}.}\label{fig:beamgas-smog}
    \end{figure}

   One of the first SMOG injections used in the LHC in 2012 with 4~TeV proton beams of
   $N \approx 8\times 10^{10}$ p per bunch is
   illustrated in Fig.~\ref{fig:beamgas-smog}, which shows the VELO pressure
    as a function
   of time (left) and the measured high level trigger (HLT) rates for beam1--gas events
   and beam2--gas events.
   Here,
   the raw Penning gauge value is displayed. A factor of 4.1 must be applied to obtain a better estimate
   of the actual neon pressure to take into account the gauge sensitivity to the gas species,
   see Ref.~\cite{Lec:Gauges}.
  %Adding a little bit of gas (here Neon) \hfill\cite{LHCb-BGI}
  %should it not be $10^{-7}$~mbar ? why $4\cdot 10^{-8}$~mbar ? see lecture \cite{Lec:Gauges}
  %Beam-gas rate increases.
  As an exercise, one can estimate the rate of beam--gas events that LHCb should have measured.
  Assuming that the LHCb HLT selected all beam--gas events with a vertex in a longitudinal range
  of about $ \Delta z=1~$m near the IP
  and assuming a flat profile of Ne pressure $P({\rm Ne}) = 1.6\times 10^{-7}$~mbar at $T= 293\UK$
  (thus $\rho_{\rm Ne} = {P({\rm Ne})} / {k_\mathrm{B} T} = 4\times 10^9~{\rm cm}^{-3}$),
  and applying the guidelines of Section~\ref{Proton beams}, using Eq.~\ref{eq:sigmapA_inel}
   to estimate the inelastic $\mathrm{p}+{\rm Ne}$ cross-section ($\sqrt{s_\mathrm{NN}} =87~$GeV, thus
  $\sigma_{{\rm inel},\mathrm{p}+\mathrm{p}}  \approx   37~{\rm mb}$, from Fig.~\ref{fig:pp-cross-section})
  \begin{equation}
     \sigma_{{\rm inel},\mathrm{p}+{\rm Ne}} \approx \sigma_{{\rm inel},\mathrm{p}+\mathrm{p}}\times 20^{0.7}
      = 37~{\rm mb} \times 8.4 = 310~{\rm  mb} ~,  % (elastics do not contribute!)
  \end{equation}
  one obtains a beam--gas rate
   \begin{equation}
     R = \sigma_{{\rm inel},\mathrm{p}+{\rm Ne}}\cdot N \cdot f\cdot \rho_{\rm Ne}\cdot \Delta z  = 110~{\rm Hz},
  \end{equation}
  a value quite close to the measured one, see Fig.~\ref{fig:beamgas-smog} (right).
  The order of magnitude is correct.
  Clearly, the devil is in the details (gauge calibration, exact cross-section, detector acceptance, efficiency),
  and a better agreement would have been a surprise.

Following the successful utilization of beam--gas events to image the beams at LHCb,
which led to an absolute luminosity calibration of less than 2\%~\cite{LHCb-BGI,LHCb-lumi},
a project was started at the LHC to build a demonstrator beam--gas vertexing   system,
based on a fibre tracker, that should allow one to measure the LHC beam transverse profile
(therefore, the transverse emittance) per bunch  within a few minutes~\cite{BGV-IPAC2017}.
This may prove to be an interesting alternative to existing beam profile measuring devices,
especially for high-energy, high-intensity beams, where standard wire scanners cannot
be applied.

\mysection{Gaseous fixed targets}

  Internal gas targets have a long past history in accelerator physics, perhaps culminating
  in the use of sophisticated polarized gas sources feeding into a cylindrical tube
  through which a beam passes (for a review, see Ref.~\cite{SteffensHaeberli}).

  In LHCb, once SMOG was made available, it was realized that interesting measurements
  could be made with gaseous targets.
  For example, the interpretation of the recent PAMELA~\cite{PAMELA-pHe} and AMS-2~\cite{AMS-pHe}
  results on the astrophysical flux
  ratio of antiprotons and protons $\Phi(\bar{\mathrm{p}})/\Phi(\mathrm{p})$ suffers from a relatively
  large uncertainty on the model predictions, owing to the lack of data on the $\bar{\mathrm{p}}$
  production cross-section $\mathrm{p}+{\rm He}\rightarrow \bar{\mathrm{p}}+X$~(for
example, see Refs. \cite{Interpret-pHe, Interpret-pHea} and the references therein).
  %Dark matter or no dark matter signal ?
  %Uncertainty in cross section: how many $\bar{p}$ produced by $p+{\rm He}$ collisions ?
   %\xxxxxx[width=0.99\textwidth]{figs/PamelaAMS.png}
  The LHC beam impinging on helium nuclei offers the opportunity for LHCb
  to perform a direct measurement of $\sigma_{\mathrm{p} +{\rm He} \rightarrow \bar{\mathrm{p}} + X}$,
  right in the relevant kinematic range.
  One of the challenges is related to the normalization of the cross-section.
  The luminosity must be measured, which in this case requires a measurement of the
  target gas density $\rho_{\rm He}(z)$ along the beam path.
  This measurement will be made in the future with the use of precisely calibrated
   Bayard--Alpert gauges, see Ref.~\cite{Lec:Gauges}.
  In the mean time, an alternative solution was found,  which again makes use of beam--gas
  scattering!
  Indeed, the proton beams also impinge on the atomic electrons of the helium atoms.
   By chance, some of the elastically scattered electrons fall inside the LHCb acceptance.
   The elastic `p + e' cross-section, is of course, identical to the elastic `e + p'
   cross-section discussed in Section~\ref{Example: elastic electron--proton scattering cross-section},
   only with a boost applied to the observer such that the electrons come to rest.

   For curiosity and as an exercise, let's calculate the boost $\beta$ needed to bring the proton to rest
   for a proton beam energy of $E_\mathrm{p} =6.5$~TeV (see Eq.~\ref{eq:boost})
   \begin{equation}
          0 =  \gamma\, (cp_z -\beta\, E_\mathrm{p})  \quad\Rightarrow \quad \beta = \frac{cp_z}{E_\mathrm{p}} %= \dyfr{\sqrt{E_p^2-\mN^2c^4}}{E_p}~.
          \quad  \Rightarrow \quad    \gamma = \frac{E_\mathrm{p}}{\mN c^2} \approx 6900~.
   \end{equation}
   Applying the same boost to the atomic electron gives the impinging electron energy
   in the proton rest frame (the binding energy is a few electronvolts and can be neglected)
   \begin{equation}
          \boo{E} =  \gamma\, m c^2  \approx 3.5~{\rm GeV},
   \end{equation}
   which is an electron beam energy similar to that used at the medium-energy electron beam facility TJNAF.

   One can derive that, in the LHCb laboratory frame, the scattered electron energy is
   deduced from its laboratory scattering angle $\theta$ by
   \begin{equation}
          E^\prime  = mc^2\, \frac{(E_\mathrm{p}+mc^2)^2+c^2\mathbf{p}_\mathrm{p}^2 \,\cos^2\theta }{(E_\mathrm{p}+mc^2)^2-c^2\mathbf{p}_\mathrm{p}^2 \,\cos^2\theta }
          \approx \frac{2\,mc^2}{\left({\mN c^2}/{E_\mathrm{p}}\right)^2+\theta^2} \quad \quad \quad (E_\mathrm{p}/c^2\gg m, \mN~\mbox{and}~\theta\ll\pi/2)
   \end{equation}
   and the four-momentum transfer squared is
   \begin{equation}
        Q^2 = 2\, m\, (E^\prime - mc^2) \approx 2\, m\, E^\prime ~ \quad \quad \quad (E^\prime\gg mc^2 ).
   \end{equation}
   At a laboratory angle of $\theta = 32~$mrad, one has $\boo{E^\prime} \approx 1~$GeV
   and $Q^2\approx 0.001~{\rm GeV}^2/c^4$. The $Q^2$ is small and therefore the
   electromagnetic form factors given in Section~\ref{Example: elastic electron--proton scattering cross-section}
   are quite precisely known.
   By measuring the rate of single electron events in non-colliding bunch slots
   (`{\be}' crossing type, see Section~\ref{Beam--gas imaging}),
   LHCb obtains a measurement of the atomic electron density $\rho_\mathrm{e}$,
   since the e + p elastic cross-section is known with precision in the relevant  $Q^2$ range.
   First results indicate that an accuracy of about 6\% can be achieved (dominated by
   systematic effects)~\cite{Graziani:2017als}.
   Assuming further that $\rho_{\rm He} = \rho_\mathrm{e} / Z_{\rm He}$, one derives the density of atoms in the LHC beam path.
   This method can and will also be applied to other target nuclei.

  LHCb has since developed a broad  physics programme based on He, Ne and Ar gas
  as a fixed target~\cite{EmilieMaurice}. Several other initiatives and ideas of
   physics measurements with beam--gas interactions at the LHC are also being proposed
   and considered. For a recent overview, see the CERN `Physics Beyond Collider' Annual Workshop~\cite{PBC-FT}.
   %and considered. For a recent overview, see the CERN `Physics Beyond Collider' forum~\cite{PBC-FT}.

%\mysection{Beam-gas interactions: other effects}
%Discuss radiation effects ?
%
%   Sorry, no time to cover:
%      radiation from beam-gas interactions (to downstream devices), see lecture \cite{Lec:Radiation}
%       and Ref.~\cite{AliceNote}
%   Apart from luminosity, very similar to radiation from collimation.
%   On can calculate that the rates from beam-gas interactions are negligible to
%   those from beam-beam collisions in the  ATLAS, CMS and LHCb  experiments.
%
%  Other exotic accelerators: muons, pions, ions, ... you dream i, were not covered.

%%%%%%%%%%%%%%%%%%%%%%%%%%%%%%%%%%%%%%%%%%%%%%%%%%%%%%%%%%%%%%

\mysection{Summary and outlook}

Beam--gas interactions in accelerators are essentially unavoidable. They are usually
a nuisance that one tries to reduce (reduced beam lifetime, radiation, background rates).
However, they can also be of practical use, allowing one to perform transverse and
longitudinal beam profile imaging or even to carry out physics experiments in storage rings.
Some simple, sometimes naive, formulae were presented that should allow one to make
order-of-magnitude estimates of the beam--gas rates in different cases, with proton, ion,
or electron beams.
For more precise estimates, the reader is encouraged to use advanced simulation codes.

\section*{Acknowledgements}

I wish to thank the organizers of this CERN Accelerator School for inviting me to give this
lecture and Antonello Di Mauro (CERN) for providing the material about the ALICE
background studies and discussing them with me.


\begin{thebibliography}{99}
% \Lecture{``Introduction to Machine Parameters'', Pedro TAVARES}{Lec:Introduction}
\bibitem{Lec:Fundamentals} E. Al Dmour, Fundamentals of vacuum technology, these proceedings.
\bibitem{Lec:Outgassing} P. Chiggiato, Materials \& properties IV: outgassing, these proceedings.
\bibitem{Lec:Dynvac} O. Malyshev, Beam induced desorption,  these proceedings.
\bibitem{Lec:Radiation} F. Cerutti, Beam induced radioactivity \& radiation hardness, these proceedings.
\bibitem{PDG} Figure downloaded from the PDG web site: \\ \url{http://pdg.lbl.gov/2014/hadronic-xsections/hadron.html} ,
exact link \\ \url{http://pdg.lbl.gov/2014/hadronic-xsections/rpp2014-pp\_pbarp\_plots.pdf}.
Not appearing in the citation: 
K.A. Olive \textit{et al.} (Particle Data Group), \textit{Chin. Phys. C}, 38, 090001 (2014). 
%\bibitem{PDG}J. Beringer \textit{et al.} (Particle Data Group), \textit{Phys. Rev. D} 86 (2012) 010001 (2012), https://doi.org/10.1103/PhysRevD.86.010001.
% ``The Review of Particle Physics (2016)'', C. Patrignani et al. (Particle Data Group), Chin. Phys. C, 40, 100001 (2016).
\bibitem{Carvalho} %%%%%%%%%%%%%%%%%%%%%%%%%
   J. Carvalho,
   \textit{Nucl. Phys. A} \textbf{725} (2003) 269, \\ \url{https://doi.org/10.1016/S0375-9474(03)01597-5}.
\bibitem{DPMJET} %%%%%%%%%%%%%%%%%%%%%%%%%

  F.W. Bopp \textit{et al.},  \textit{Phys. Rev. C} \textbf{77} (2008) 014904, \\ \url{https://doi.org/10.1103/PhysRevC.77.014904}.
\bibitem{Epos} %%%%%%%%%%%%%%%%%%%%%%%%%
    B. Guiot and K. Werner, \textit{J. Phys. Conf. Series} \textbf{589-1} (2015) 012008, \\ \url{http://stacks.iop.org/1742-6596/589/i=1/a=012008}.
\bibitem{QGSJETII-04} %%%%%%%%%%%%%%%%%%%%%%%%%
   S. Ostapchenko,
   \textit{Phys. Rev. D} \textbf{81} (2010) 114028, \\ \url{https://doi.org/10.1103/PhysRevD.81.114028}.
   \bibitem{HIJING-1.383} %%%%%%%%%%%%%%%%%%%%%%%%%
    X.-N. Wang and M. Gyulassy, \textit{Phys. Rev. D} \textbf{44} (1991) 3501, \\ \url{https://doi.org/10.1103/PhysRevD.44.3501}.
 %\\ \url{http://ntc0.lbl.gov/~xnwang/hijing/}.
%\bibitem{EPOS-LHC} K. Werner, F.-M. Liu, and T. Pierog, ``Parton ladder splitting and the rapidity dependence
%of transverse momentum spectra in deuteron-gold collisions at RHIC'' Phys. Rev. C 74
%(2006) 044902, doi:10.1103/PhysRevC.74.044902, arXiv:hep-ph/0506232,
\bibitem{FLUKA} %%%%%%%%%%%%%%%%%%%%%%%%%
       T.T. B\"ohlen \textit{et al.}, \textit{Nucl. Data Sheets} \textbf{120}  (2014) 211, \\ \url{https://doi.org/10.1016/j.nds.2014.07.049},
  \\ \url{http://www.fluka.org/fluka.php}.
\bibitem{Geant} J. Allison \textit{et al.},
   \textit{Nucl. Instrum. Methods Phys. Res. A} \textbf{835} (2016) 186, \\ \url{https://doi.org/10.1016/j.nima.2016.06.125},
 \\ \url{http://geant4.cern.ch/}.
\bibitem{Pythia} %%%%%%%%%%%%%%%%%%%%%%%%%
     T. Sj\"ostrand \textit{et al.}, \textit{Comput. Phys. Comm.} \textbf{178} (2008) 852, \\ \url{https://doi.org/10.1016/j.cpc.2008.01.036},
\\ \url{http://home.thep.lu.se/~torbjorn/pythia81html/Welcome.html}.
\bibitem{LowEnergy}    J.R. Letaw \textit{et al.},  \textit{Astrophys. J. Suppl. Series} \textbf{51} (1983) 271, \\ \url{https://doi.org/10.1086/190849}.
    %\\ \url{http://adsbit.harvard.edu//full/1983ApJS...51..271L/0000271.000.html}.
\bibitem{Hoang}  %%%%%%%%%%%%%%%%%%%%%%%%%
     T.F. Hoang \textit{et al.}, \textit{Z. Phys C} \textbf{29} (1985) 611, \url{https://doi.org/10.1007/BF01560296}.
   %\href{http://link.springer.com/article/10.1007/BF01560296}{http://link.springer.com/article/10.1007/BF01560296}.
\bibitem{Donnelly} %%%%%%%%%%%%%%%%%%%%%%%%%
   T.W. Donnelly and J.D. Walecka,
    \textit{Ann. Rev. Nucl. Sci.} \textbf{25} (1975) 329, \\ \url{https://doi.org/10.1146/annurev.ns.25.120175.001553}.
\bibitem{Tsai} %%%%%%%%%%%%%%%%%%%%%%%%%
    Y.-S. Tsai,
     \textit{Rev. Mod. Phys.} \textbf{46}  (1974) 81.  Erratum, \textit{Rev. Mod. Phys.} \textbf{49} (1977) 421, \\ \url{https://doi.org/10.1103/RevModPhys.46.815}.
\bibitem{ep-elastic} %%%%%%%%%%%%%%%%%%%%%%%%%
   D.H. Perkins, \textit{Introduction to High Energy Physics}, 3rd ed.
   (Addison-Wesley, Boston, 1987).
\bibitem{ep-emff} %%%%%%%%%%%%%%%%%%%%%%%%%
   C. Perdrisat and V. Punjabi, \textit{Scholarpedia} {\bf 5} (2010) 10204, 
   \\ \url{https://doi.org/10.4249/scholarpedia.10204}, revision \#143981,
   \\ \url{http://www.scholarpedia.org/article/Nucleon\textunderscore Form\textunderscore factors}.

\bibitem{HerrMuratori}  %%%%%%%%%%%%%%%%%%%%%%%%%

   W. Herr and B. Muratori, Concept of luminosity,    CAS---CERN Accelerator School: Intermediate Course on Accelerator Physics, Zeuthen, 2003, p. 361, \\ \url{https://doi.org/10.5170/CERN-2006-002}.
\bibitem{ATLAS-BIB} ATLAS Collaboration, \textit{J. Instrum.} \textbf{8} (2013) P07004, \\ \url{http://iopscience.iop.org/1748-0221/8/07/P07004}.
\bibitem{ATLAS-BIB2}  %%%%%%%%%%%%%%%%%%%%%%%%%
     G.~Aad \textit{et al.} \textit{J. Instrum.} \textbf{11} (2016) P05013, \\ \url{http://stacks.iop.org/1748-0221/11/i=05/a=P05013}.
\bibitem{Gibson}  %%%%%%%%%%%%%%%%%%%%%%%%%
  S.~Gibson \textit{et al.}, Beam-gas background observations at LHC,
Proc. 8th Int. Particle Accelerator Conf. (IPAC'17),
   Copenhagen, 2017, ATL-DAPR-PROC-2017-002, paper TUPVA032, p. 2129, \\ \url{https://doi.org/10.18429/JACoW-IPAC2017-TUPVA032}.

\bibitem{ALICE} %%%%%%%%%%%%%%%%%%%%%%%%%
  ALICE Collaboration, \textit{Int. J. Mod. Phys. A} \textbf{29} (2014) 1430044, \\ \url{https://doi.org/10.1142/S0217751X14300440}.
  %\\ \url{http://www.worldscientific.com/doi/abs/10.1142/S0217751X14300440}.
\bibitem{ALICE-TPC} %%%%%%%%%%%%%%%%%%%%%%%%%
    Alme  \textit{et  al.},  \textit{Nucl. Instrum. Methods Phys. Res. A} \textbf{622} (2010) 316, \\ \url{https://doi.org/10.1016/j.nima.2010.04.042}.
\bibitem{AliceV0} %%%%%%%%%%%%%%%%%%%%%%%%%
 ALICE Collaboration,
   \textit{J. Instrum.} \textbf{8} (2013) P10016, \\ \url{https://doi.org/10.1088/1748-0221/8/10/P10016}.  %\href{http://stacks.iop.org/1748-0221/8/i=10/a=P10016}{http://stacks.iop.org/1748-0221/8/i=10/a=P10016}.
%\bibitem{AliceNote} %%%%%%%%%%%%%%%%%%%%%%%%%
%   ``Radiation Dose and Fluence in ALICE after LS2'',
%    A. Alici, A. Di Mauro, W. Riegler and A.Tauro, public ALICE note, to be published (CERN, Geneva).
%\bibitem{KEK-BEAST} %%%%%%%%%%%%%%%%%%%%%%%%%
%    KEK-BEAST ref xxxxxxxxxxxx
\bibitem{LHCb} LHCb Collaboration, \textit{Int. J. Mod. Phys. A} \textbf{30} (2015) 1530022, \\ \url{https://doi.org/10.1142/S0217751X15300227}.
      %\href{http://www.worldscientific.com/doi/abs/10.1142/S0217751X15300227}{http://www.worldscientific.com/doi/abs/10.1142/S0217751X15300227}.
\bibitem{LHCb-BGI}  %%%%%%%%%%%%%%%%%%%%%%%%%
  C. Barschel, Ph.D. thesis, Rheinisch-Westf\"alische Technische Hochschule,
2013,  CERN-THESIS-2013-301, \\ \url{https://cds.cern.ch/record/1693671}.
\bibitem{LHCb-lumi}  %%%%%%%%%%%%%%%%%%%%%%%%%
  LHCb collaboration,
  \textit{J. Instrum.}  \textbf{9} (2014) P12005, \\ \url{https://doi.org/10.1088/1748-0221/9/12/P12005},
     \\ \url{http://stacks.iop.org/1748-0221/9/i=12/a=P12005}.
\bibitem{VELO} %%%%%%%%%%%%%%%%%%%%%%%%%
    R. Aaij \textit{et al.}, \textit{J. Instrum.} 9 (2014) P09007, \\ \url{https://doi.org/10.1088/1748-0221/9/09/P09007}.   % \href{http://stacks.iop.org/1748-0221/9/i=09/a=P09007}{http://stacks.iop.org/1748-0221/9/i=09/a=P09007}.
%\bibitem{Lec:ThinFilm} %%%%%%%%%%%%%%%%%%%%%%%%%
%    ``Thin-Film Coating'' Pedro Costa Pinto,  in these Proceedings.
\bibitem{DCCT} C. Barschel \textit{et al.},   %%%%%%%%%%%%%%%%%%%%%%%%%
    Results of the LHC DCCT calibration studies, CERN-ATS-Note-2012-026 PERF
(CERN, Geneva, 2012), \\ \url{https://cds.cern.ch/record/1425904}.
\bibitem{BCNWG3}  %%%%%%%%%%%%%%%%%%%%%%%%%
  G. Anders  \textit{et al.}, Study of the relative LHC bunch populations for luminosity calibration,
   CERN-ATS-Note-2012-028 PERF, BCNWG Note 3 (CERN, Geneva, 2012), \\ \url{https://cds.cern.ch/record/1427726}.
\bibitem{Lec:Getters}  %%%%%%%%%%%%%%%%%%%%%%%%%
  E. Maccallini, Getter pumps, these proceedings.
\bibitem{Lec:Gauges}  %%%%%%%%%%%%%%%%%%%%%%%%%
   K. Jousten, Vacuum gauges I \& II, these proceedings.
\bibitem{BGV-IPAC2017}  %%%%%%%%%%%%%%%%%%%%%%%%%
  A. Alexopoulos \textit{et al.}, First LHC transverse beam size measurements with the beam gas vertex detector,
Proc. 8th Int. Particle Accelerator Conf. (IPAC'17),
   Copenhagen,  2017, paper TUOAB1, p. 1240, \url{https://doi.org/10.18429/JACoW-IPAC2017-TUOAB1}.
\bibitem{SteffensHaeberli}  %%%%%%%%%%%%%%%%%%%%%%%%%
  E. Steffens and W. Haeberli,
  \textit{Rep. Prog. Phys.} \textbf{66} (2003) 1887, \\ \url{http://stacks.iop.org/0034-4885/66/i=11/a=R02}.
\bibitem{PAMELA-pHe}  %%%%%%%%%%%%%%%%%%%%%%%%%
   O. Adriani \textit{et al.}, \textit{Science} \textbf{332} (2011) 69, \\ \url{https://doi.org/10.1126/science.1199172}.
\bibitem{AMS-pHe} %%%%%%%%%%%%%%%%%%%%%%%%%
   M. Aguilar \textit{et al.} (AMS Collaboration), \textit{Phys. Rev. Lett.} \textbf{117} (2016) 091103, \\ \url{https://doi.org/10.1103/PhysRevLett.117.091103}.
\bibitem{Interpret-pHe} %%%%%%%%%%%%%%%%%%%%%%%%%
   M.-Y. Cui \textit{et al.}, \textit{Phys. Rev. Lett.} \textbf{118} (2017) 191101, \\ \url{https://doi.org/10.1103/PhysRevLett.118.191101}.
%``The cosmic ray antiproton background for AMS-02'', R. Kappl and M.W. Winkler, JCAP09 (2014) 051.

\bibitem{Interpret-pHea} %%%%%%%%%%%%%%%%%%%%%%%%%
   A. Cuoco \textit{et al.}, \textit{Phys. Rev. Lett.} \textbf{118} (2017) 191102, % M. Kr\"amer and M. Korsmeier, Phys. Rev. Lett. 118, (2017) 191102,
     \\ \url{https://doi.org/10.1103/PhysRevLett.118.191102}. %``The cosmic ray antiproton background for AMS-02'', R. Kappl and M.W. Winkler, JCAP09 (2014) 051.

\bibitem{Graziani:2017als}  %%%%%%%%%%%%%%%%%%%%%%%%%
 G.~Graziani (on behalf of LHCb Collaboration), Fixed target measurements at LHCb for cosmic rays physics, Proc. 52nd Rencontres de Moriond on EW Interactions and
   Unified Theories (Moriond EW 2017), La Thuile, 2017,
   arXiv:1705.05438 [hep-ex].
\bibitem{EmilieMaurice}  %%%%%%%%%%%%%%%%%%%%%%%%%
  E. Maurice (on behalf of LHCb Collaboration),
Fixed-target physics at LHCb, Proc. 5th Conf.  Large Hadron Collider Physics 2017, Shanghai, 2017,
        arXiv:1708.05184 [hep-ex].
\bibitem{PBC-FT}  %%%%%%%%%%%%%%%%%%%%%%%%%
% http://pbc.web.cern.ch/.
  Physics Beyond Collider Annual Workshop, 21-22 November 2017, CERN, \\ \url{https://indico.cern.ch/event/644287}.
  In particular, G. Graziani's talk on LHCb as a fixed-target experiment, %LHCb-TALK-2017-384, http://cds.cern.ch/record/2294176.
                 P. di Nezza's talk on Polarized fixed target at LHC,
                 L.M. Massacrier's talk on ALICE fixed target and
                 J.-P. Lansberg's talk on AFTER.

%\bibitem{Alice-bkg} %%%%%%%%%%%%%%%%%%%%%%%%%
%  ``ALICE vacuum requirements and TDI-related background issues in Run 1+2'',
%  A. Di Mauro, in TDIS Internal Review, 1 Dec. 2016,  CERN,
%  \href{https://indico.cern.ch/event/579995/timetable}{https://indico.cern.ch/event/579995/timetable}.
%\bibitem{LHCb-pHe}  %%%%%%%%%%%%%%%%%%%%%%%%%
%  ``Measurement of antiproton production in pHe collisions at $\sqrt{s_{NN}}=110~$GeV'',
%   CERN-LHCb-CONF-2017-002, The LHCb Collaboration,
%   52nd Rencontres de Moriond on Electroweak Interactions and Unified Theories, La Thuile, Italy, 18 - 25 Mar 2017,
%   \href{https://cds.cern.ch/record/2260835}{https://cds.cern.ch/record/2260835}.
%\bibitem{Sixtrack} %%%%%%%%%%%%%%%%%%%%%%%%%
%  SixTrack - 6D Tracking Code,
%  \href{http://sixtrack.web.cern.ch/SixTrack/}{http://sixtrack.web.cern.ch/SixTrack/}.
%%%% from CMS p-Pb paper
\end{thebibliography}
\end{document}